\documentclass[twocolumn,floatfix,superscriptaddress,amsmath,showpacs,showkeys,aps,prb,eqsecnum]{revtex4-1}
\usepackage[utf8]{inputenc}
\usepackage{caption}
\usepackage{graphicx}
\usepackage{bm}
\usepackage[normalem]{ulem}
\usepackage[usenames]{color}
\usepackage{times}
\usepackage{verbatim}
\usepackage{xcolor}
\usepackage{hyperref}
\usepackage{subfigure}
\newcommand{\sgn}{\mathop{\mathrm{sgn}}}

\def\C60{A$_x$C$_{60}$}

\def\SROtwo{ Sr$_{3}$Ru$_{2}$O$_{7}$}

\def\HgCu3{HgCa$_2$Cu$_3$O$_{8+y}$}
\def\HgCu4{HgBa$_2$Ca$_3$Cu$_4$O$_{10+y}$}
\def\TlCu{Tl$_2$Ba$_2$CuO$_{6+\delta}$}
\def\TlCu3{Tl$_2$Ba$_2$Ca$_2$Cu$_3$O$_{10+y}$}
\def\TlCu4{Tl$_2$Ba$_2$Ca$_3$Cu$_4$O$_{12+y}$}

\def\BiCu3{Bi$_2$Sr$_2$Ca$_{2}$Cu$_3$O$_y$}

\def\8LSCO{La$_{1.88}$Sr$_{.12}$CuO$_4$}
\def\110LNSCO{La$_{1.5}$Nd$_{0.4}$Sr$_{0.1}$CuO$_{4}$}
\def\stage4LCO{La$_{2}$CuO$_{4+\delta}$}
\def\Y248{YBa$_2$Cu$_4$O$_8$}

\def\NbSe2{NbSe$_2$}
\def\TaSe2{TaSe$_2$}
\def\TiSe2{TiSe$_2$}
\def\NaCoOH2O{Na$_{0.3}$CoO$_{2y}$H$_2$O}
\def\MgB2{MgB${}_2$}

\def\URu2Si2{URu$_2$Si$_2$}

\def\Ba122{Ba(Fe$_{1-x}$Co$_x$)$_2$As$_2$}

\def\hts{high temperature superconductors}

\begin{document}
\title{Bosonization of  Fermi liquids in a weak  magnetic field}
\author{Daniel G. Barci}
\affiliation{Departamento de F{\'\i}sica Te\'orica,
Universidade do Estado do Rio de Janeiro, Rua S\~ao Francisco Xavier 524, 20550-013  
Rio de Janeiro, Brazil}
\author{Eduardo Fradkin}
\affiliation{Department of Physics and Institute for Condensed Matter Theory, University of Illinois 
at Urbana-Champaign, 1110 W. Green Street, Urbana, Illinois 61801-3080, U.S.A.}
\author{Leonardo Ribeiro}
\affiliation{Departamento de F{\'\i}sica Te\'orica,
Universidade do Estado do Rio de Janeiro, Rua S\~ao Francisco Xavier 524, 20550-013  
Rio de Janeiro, Brazil}

\date{\today}

\begin{abstract}
Novel controlled non-perturbative techniques are a must in the study of strongly correlated systems, especially near quantum criticality. 
One of these techniques, bosonization, has been extensively used to understand one-dimensional,  as well as higher dimensional electronic systems at finite density. In this paper, we generalize the theory of two-dimensional bosonization of Fermi liquids, in the presence of  a homogeneous weak magnetic field perpendicular to the plane.   Here, we extend the formalism of bosonization to treat  free spinless fermions at finite density in a uniform magnetic field. 
We show that particle-hole fluctuations of a Fermi surface satisfy a {\em covariant Schwinger algebra}, allowing to express a fermionic theory with forward scattering interactions as a quadratic bosonic theory representing the quantum fluctuations of the Fermi surface.
By means of a coherent-state path integral  formalism we  compute the fermion propagator as well as particle-hole bosonic correlations functions. We  analyze the presence of de Haas-van Alphen oscillations and   show how the quantum oscillations of the orbital magnetization, the Lifshitz-Kosevich theory,  are  obtained by means of the bosonized theory.   
We  also study the effects of forward scattering interactions. In particular, we obtain oscillatory corrections to the Landau zero sound collective mode.  
\end{abstract}

\maketitle

\section{Introduction}
\label{Sec:Introduction}

Consistent advances in our understanding of strongly correlated fermionic systems depend on the 
development of controlled  non-perturbative computation  techniques.
Important questions, such as the  fate of interacting systems of fermions at finite density in a quantum critical regime,\cite{Hertz1976,Millis1993} are still waiting to be answered. This is particularly pressing given the  ``non-Fermi-liquid'' behaviors seen in many systems, notably in {\hts} in their ``strange metal'' regime, and many others systems, such as {\SROtwo}, and heavy-fermion materials near quantum critical points.

At  the theoretical level, bosonization  has played an important role 
in developing accurate low-energy theories in one and quasi-dimensional systems.\cite{Lieb-1965,Luther-Emery-1974,Mandelstam-1975,Coleman-1975,Emery-1979,Haldane-1981}
The applicability of its higher dimensional generalization of this approach to describe the  quantum critical behaviour of fermionic systems at finite density (i.e. with a Fermi surface) is still under study. Recent significant progress has been made in the development of bosonization dualities for theories of relativistic fermions in 2+1 dimensions.\cite{Son-2015,Karch-2016,Seiberg-2016}

The first step in the development of higher dimensional bosonization for Fermi fluids was made by Luther in the eighties.\cite{Luther1979} Intense activity in the 1990s in this area lead to the description of  the Fermi liquid fixed point.\cite{CastroNeto-1993,CastroNeto-1994,CastroNeto-1995,houghton-1993,houghton-1994,haldane-1994,houghton-2000} 
The essential idea  is that most relevant fermionic properties at low energies can be described by the dynamics of particle-hole excitations  near the Fermi surface. In this sense, the Fermi surface can be considered as a quantum mechanical extended object, a membrane in momentum space, with its own quantum dynamics. 
Renormalization group methods for  fermionic systems at finite density\cite{Shankar1994} found that two-body forward scattering interactions, parametrized by Landau parameters, are marginally irrelevant at the Fermi liquid fixed point, consistent with the description provided by the Landau theory of Fermi liquids. \cite{nozieres-1999}

Higher dimensional bosonization turned out to be a powerful tool to  study  metallic systems  away form the  Fermi liquid regime. Different phase transitions, for instance driven by Pomeranchuk instabilities, were studied using this approach,\cite{BaOx2003} in particular  the isotropic-nematic quantum phase transition.\cite{OgKiFr2001} It was shown that  the nematic quantum critical point and as the nematic phase itself  are dominated by an overdamped low energy mode, a Goldstone mode with dynamical critical exponent $z=3$.\cite{Lawler2006} In these regimes, although the correlators of the order parameter scale both in frequency and momentum (with a certain dynamic critical exponent $z$), the fermionic correlators scale only in their frequency dependence, i.e. exhibit a form of  ``local quantum criticality.''\cite{lawler-2007}

A powerful experimental tool used to study the properties of Fermi systems, and particularly their Fermi surface, are  quantum oscillation experiments,  in particular  the de Haas-van Alphen effect (dHvA).  In this case, the system is placed in a strong enough magnetic field that overcomes temperature effects, however weak enough to be far away from the Landau level quantization regime. In these conditions, thermodynamic observables such as the magnetization and susceptibility, and the heat capacity, exhibit oscillations as the magnetic field is varied. The Lifshitz-Kosevich (LK) theory, \cite{kosevich-1956-1,kosevich-1956-2} a semiclassical theory of free fermions in a magnetic field,  relates  the oscillations with geometric properties of the Fermi surface.
This procedure has been used for a long time to study normal metals which are well described by the Landau theory of the Fermi liquid. 

However, more recently, quantum oscillations have been observed in strongly correlated materials. The Lifshitz-Kosevich theory is nowadays widely used  to determine Fermi surface properties in strongly correlated systems,
\cite{Doiron-Leyraud2007,Sebastian2008,Vignolle2008,Yelland2008,Jaudet2008,Audouard2009,Barisic2013}, even  near a quantum critical point\cite{Borzi2007}. However, the validity of the LK theory is questionable in such extreme regimes where Fermi liquid theory is known to fail.
Thus, understanding quantum oscillations in a strong coupling regime and, particularly, near a quantum critical point is an important and open problem.\cite{senthil2009,Hartnoll2010,ChWeBeSe-2018}
Generally, a magnetic field opens a gap in the fermionic spectrum and is a relevant perturbation that moves the system away form the critical point. How the system scales when the magnetic field approach zero at a Pomeranchuk instability\cite{BaRe2013} is a key question to understand the more general problem of quantum oscillations. 

The main purpose of this paper is to extend the bosonization procedure in the presence of a weak homogeneous magnetic field. 
Here we will focus on the case of two-dimensional systems. The extension to higher dimensions is straightforward. 
In this paper, we  develop a bosonization approach to a two dimensional spinless fermionic system, in the presence of a magnetic field $B$ perpendicular to the plane. 
Previous attempts to bosonize  a two-dimensional Fermi gas in a weak magnetic field were done in the Landau level basis\cite{Caldeira-1997}, {\em i. \ e.\ }, by first projecting the fermionic fields into Landau levels. This technique was successfully applied to solve some problems in which static localized interactions produce transitions between, almost non-interacting, Landau levels.\cite{Caldeira-1998}
Here,  we follow a different approach, keeping the formalism as close as possible to the conventional bosonization of Fermi liquids.  With this tool, we expect to be able to compute correlations in the dHvA regime in strongly coupled fermionic systems.  

As usual in bosonization, we consider a high density regime in which the Fermi surface can be taken to be locally flat. Assuming that the low energy physics is driven by small fluctuations of the shape of the Fermi surface,\cite{CastroNeto-1993} we project the Hamiltonian into a narrow shell of states, divided into patches, around the Fermi surface. The patches are small enough to allow the linearization of  the dispersion relation, but big enough to contain a large number of Landau levels. 

In this regime, it is possible to define local Fermi surface fluctuations in terms of densities particle-hole excitations and show that the corresponding operators satisfy a {\em covariant Schwinger algebra}. This algebra is at the root of the bosonization approach, see Eq. (\ref{eq:Schwinger-algebra}) below.
Within this scheme,  the fermion operator is defined as an exponential of a coherent superposition of bosonic fields, and show that this construction is  consistent with the basic symmetries of the low energy theory.  Importantly, the fermion operator in each patch depends on a superposition of bosonic excitations all around the Fermi surface.   

We also present an extension of the  coherent-state path integral formalism\cite{CastroNeto-1993} to describe a system in the presence of a magnetic field, suitable for the computation of correlation functions. The semiclassical approximation of this theory leads to an  equation of motion that  is a particular case of the Landau-Silin\cite{nozieres-1999} equation in the Fermi liquid theory. Using these methods we  computed the bosonic correlation functions necessary to characterize the  linear response of the system.  This theory allowed us  to study the emergence of the dHvA oscillations from a  bosonized approach.
We  also discuss the effects of forward scattering interactions within this formalism. In particular we  compute the collective mode spectrum, finding damped oscillatory corrections to  Landau zero sound. This is a manifestation of quantum oscillations in a Fermi liquid non-equilibrium property.
 
The paper is organized as follows: in Section \ref{Sec:Fermi-gas} we compute an asymptotic form of the two-dimensional propagator of a  Fermi gas in the regime of weak magnetic fields.  In \S \ref{Sec:Bosonization} we present the main results of the bosonization approach. Then, we study symmetries of the low energy Hamiltonian in Section  \ref{Sec:Symmeties}.  In Section  \ref{Sec:FermionOperator} we construct the fermionic operators and compute the fermion propagator by using the bosonized theory.  In \S\ref{Sec:BosDyn} we present a path integral representation of Fermi surface fluctuations and compute bosonic correlation functions. In Section \ref{Sec:dHVA} we show how the dHvA oscillations can be computed from the bosonized action.  In \S \ref{Sec:Interactions} we discuss forward scattering interactions, showing an explicit calculation of collective modes and their dependence on the magnetic field. Finally we discuss our results in Section \ref{Sec:Conclusions} . Details of the calculations are presented in several appendices. 

\section{Fermi gas in a weak homogeneous magnetic field}
\label{Sec:Fermi-gas}
In this section we present a convenient representation of the asymptotic fermion propagator of a Fermi gas at high density  in a weak magnetic field.
The Fermi gas in a magnetic field is a very well known system. The effect of the magnetic field is, essentially, to discretize the fermionic spectrum into highly degenerated equally spaced Landau Levels. The gaps and the degeneracy grow with the strength of the magnetic field. Then, for sufficiently high magnetic field, a finite density of electrons will occupy just the first Landau level. In the presence of interactions, this regime leads to the integer and fractional quantum Hall fluids.  In this work, we are interested, instead, in the regime of low magnetic fields, where there is a huge number of  filled Landau levels in a small energy interval around the Fermi energy. 
The motivation is two-fold. We will first use conventional methods to derive an explicit expression for the fermion propagator computed directly, in order to compare it with the results obtained by multidimensional bosonization. On the other hand, the expression for the occupation number, obtained from the propagator,  will be later used  to derive the bosonic algebra in the bosonization procedure.  The exact propagator of a Fermi gas is well known and was obtained by several methods, see e.g. Refs.
[\onlinecite{dodonov1975,Isihara1979-1,Isihara1979-2}]. In this paper we present an asymptotic form of the fermion propagator, especially useful in the weak magnetic field regime we will be interested in.

We begin by considering a  two-dimensional  degenerated gas of spinless fermions  in a perpendicular homogeneous magnetic field. The Hamiltonian is, 
\begin{equation}
H=\int d^2x\;  \frac{1}{2 m}   \left|  \left(   \bm{\nabla}+ie \bm{A}({\bm x})  \right)\psi({\bm x})\right|^2\; , 
\label{eq:Hamiltonian}
\end{equation}
where $\psi^{\dagger}({\bm x})$ and $\psi({\bm x})$ are  fermionic creation and annihilation  operators at position ${\bm x}$, satisfying, 
\begin{equation}
\left\{\psi^{\dagger}({\bm x}),\psi({\bm x}')\right\}=i\delta^2({\bm x}-{\bm x}')\; .
\label{eq:anticomutation-x}
\end{equation}
$m$ is the effective electron mass, $e$ is the electron electric charge and the homogeneous magnetic field is given by ${\bm \nabla}\times {\bm A}=B{\hat {\bm z}}$. In two dimensions, the magnetic field $B$ is a pseudo-scalar.

The fermion propagator is defined as 
\begin{equation}
i G_F(\bm{x},\bm{y}, t,t')=\langle  FS
| {\cal T}\psi(x,t)\psi^{\dagger}(y,t')|FS\rangle. \; , 
\end{equation}
where ${\cal T}$ means time ordered product, and the filled Fermi sea ground state  $|FS\rangle$ is built by the applications of the fermion creation  operators $c^{\dagger}_{n, k}$ on the vacuum state, filling up the single particle states up to the Fermi energy $\epsilon_F$:   
\[
|FS\rangle=\prod_{n,k}  c^{\dagger}_{n,k} |0> \; .
\]
Here,  $\{n,k\}$ are the quantum numbers of the filled Landau levels. The energy eigenvalues are $E_{n,k}=\hbar \omega_c(n+1/2)$ with the cyclotron frequency $\omega_c=eB/m$, in units in which $\hbar =1$.
At zero magnetic  field, the ground state $|FS\rangle$ is a completely filled circular Fermi surface of radius $k_F$, and chemical potential $\mu=\epsilon_F=k_F^2/2m$. 
At zero or very low temperatures, the dynamics is governed by the states laying in the  small momentum  shell  
$k_F-\lambda/2< |{\bm k}|<k_F+\lambda/2$. Equivalently, the energy shell can be written as  $\epsilon_F-\Delta\epsilon< E <\epsilon_F+\Delta\epsilon$. Since $\lambda\ll k_F$, the single-particle energy dispersion can be linearized around $k_F$, and   
$\Delta\epsilon=v_F\lambda/2$, where $v_F$ is the Fermi velocity. From now on, we use  units in which $c=1$, $e=1$, $\hbar=1$.

The number of filled Landau levels contained in the momentum shell below the Fermi surface  is $N_L=\Delta\epsilon/\omega_c=\lambda\ell_B^2k_F/2$, where we have introduced the magnetic length $\ell_B^2=1/B=v_F/(k_F\omega_c)$. Even in the limit $\lambda\ll k_F$, we choose the magnetic field  in such a way that  there is still a macroscopic number of filled Landau levels, $N_L\gg 1$, in the energy shell. This condition can be stated as $\omega_c\ll \lambda v_F$, or equivalently, $\ell_B^2\gg 1/(\lambda k_F)$.  

In Appendix \ref{Ap:Propagator} we  compute the fermion propagator in this regime, and  obtain the explicit expression 
\begin{equation}
i G_F({\bm x}, {\bm y}, r_0)=\frac{ik_F}{2\pi} e^{-i\epsilon_F r_0}e^{i\theta_B}  \int_0^{2\pi} d\varphi\; \sum_{\ell=-\infty}^{+\infty}G_{\ell,\varphi}(r, r_0)\; , 
\label{eq:Propagator}
\end{equation}
where  $r=|{\bm x}-{\bm y}|$ is the spacial separation, and $r_0=t_x-t_y$ is the time difference.  
This propagator depends on the choice of gauge. This dependence  is contained in the phase $\theta_B$.
For instance, in the Landau gauge,  $A_1=-B x_2$, $A_2=0$, where we have chosen a cartesian coordinates system ${\bm x}=(x_1,x_2)$, ${\bm y}=(y_1,y_2)$), the phase $\theta_B$ is 
\begin{equation}
\theta_B({\bm x}, {\bm y})=
\frac{1}{2\ell_B^2} (x_1-y_1)(x_2+y_2)  \; .
\label{eq:ThetaLandau}
\end{equation}
Instead, in the the symmetric gauge $A_1=-B x_2/2$, $A_2=B x_1/2$,  the phase is
\begin{equation}
\theta_B({\bm x}, {\bm y})=
\frac{1}{2\ell_B^2} (x_1y_2-y_1x_2) \; .
\label{eq:ThetaSymmetric}
\end{equation}
The rest of the expression for the propagator is gauge-invariant. 

In Eq. \eqref{eq:Propagator} we used the following notation
\begin{equation}
G_{\ell,\varphi}(r, r_0)=\frac{e^{ik_F r\cos\varphi}}
{r\cos\varphi-v_F r_0+2\pi\ell \ell_B^2 k_F + i\alpha\sgn(r_0)}\; .
\label{eq:Propagator-ellvarphi}
\end{equation}
where $\varphi$ denotes the direction normal to the Fermi surface, i.e. $\cos\varphi={\bm r}\cdot {\bm k}_F/rk_F$, and $\ell$ is an integer. Eqs. \eqref{eq:Propagator} and \eqref{eq:Propagator-ellvarphi} are the main results of this section. 

In order to check the $B\to 0$ limit, 
it is convenient to sum over $\ell$ in order to have a closed form for the propagator. We obtain, 
\begin{align}
&i G_F(\bm{x}, \bm{y}, r_0)=\frac{i k_F}{2\pi} e^{-i\epsilon_F r_0}e^{i\theta_B} \int d\varphi\;   e^{ik_F r\cos\varphi} 
\nonumber \\
&\times
\frac{1}{2\ell_B^2 k_F}\cot\Big(\frac{1}{2\ell_B^2 k_F} \left(r\cos\varphi-v_F r_0+i\alpha\sgn(r_0) \right) \Big) \; .
\end{align}
Expanding the cotangent in powers of $1/\ell_Bk_F$ we obtain, 
\begin{align}
&i G_F({\bm x}, {\bm y}, r_0)=\frac{i k_F}{2\pi} e^{-i\epsilon_F r_0}e^{i\theta_B}\nonumber\\
&\times  \int d\varphi\;   e^{ik_F r\cos\varphi} 
\left\{
\frac{1}{r\cos\varphi-v_F r_0+i\alpha\sgn(r_0)}\right. \nonumber \\
&~~~-\left.\frac{1}{12\ell_B^4 k_F^2} \left(r\cos\varphi-v_F r_0\right) \right\} \; , 
\label{eq:expansion}
\end{align}
where the first term is the asymptotic Fermi liquid propagator, first computed by Luther,\cite{Luther1979} and the second term is the leading order correction in the small magnetic field expansion.  

Returning to Eq.\eqref{eq:Propagator-ellvarphi}, and after Fourier transforming its expression,  we find the asymptotic propagator in momentum and frequency space, 
\begin{align}
\tilde G_{\varphi}({\bm q}, \omega)&=\frac{1}{\omega-{\bm v}_F\cdot {\bm q}+iv_F\lambda \sgn(\omega)} \nonumber \\
&\times\sum_{\ell=0}^{\infty}
e^{-2\pi\ell N_L} \cos\left(2\pi\ell \frac{{\bm v}_F\cdot {\bm q}}{\omega_c}\right). 
\label{eq:Gvarphi-q}
\end{align}
Thus, the propagator oscillates as a function the energy of the particle-hole pair ${\bm v}_F\cdot {\bm q}$ with period $\omega_c/\ell$, damped by the number of filled of Landau level $N_L$. In the limit of $\omega_c\to 0$, $N_L\to \infty$ and only  $\ell=0$ contributes, recovering the well known Fermi gas result.

\subsection{Occupation number}
We will now discuss  the structure of the occupation number to check the consistency of the asymptotic propagator given by Eq. \eqref{eq:Propagator}.
It will needed below  to develop the bosonization technique. 
The occupation number, $n(\bm{k})$, is given by the equal-time expression
\begin{equation}
n({\bm k})=\int d^2x \; iG_F({\bm x},0,0) e^{i{\bm k}\cdot {\bm x}}   \; .
\label{eq:n-def}
\end{equation}
Replacing Eq. \eqref{eq:Propagator} into Eq. \eqref{eq:n-def}, we  write
\begin{equation}
n({\bm k})=\int_0^{2\pi}  \frac{d\varphi}{2\pi}\;   n_{\varphi}({\bm k})  \; , 
\label{eq:n-Int-varphi}
\end{equation}
with 
\begin{equation}
n_\varphi({\bm k})=ik_F  \sum_{\ell=-\infty}^{+\infty}  \int d^2x\;  e^{i\theta_B({\bm x})} \; G_{\ell,\varphi}({\bm x}, 0) \; e^{-i{\bm k}\cdot {\bm x}} 
\; . 
\label{eq:n1}
\end{equation} 
To compute the integrals we define a local coordinate system for fixed direction $\varphi$. Since  ${\bm k}_F=k_F (\cos\varphi,\sin\varphi)$, we  define a local frame of two orthonormal  vectors 
$\hat {\bm n}_\varphi$ and $\hat {\bm t}_\varphi$, respectively defining the local normal and tangential directions to the Fermi surface.  In this basis, position and momentum are decomposed as ${\bm x}=x_n \hat {\bm n}_\varphi+x_t \hat {\bm t}_\varphi$,  and ${\bm k}=k_n \hat {\bm n}_\varphi+k_t \hat {\bm t}_\varphi$. Thus, in the local frame,  Eq. \eqref{eq:n1} reads, 
\begin{equation}
n_\varphi({\bm k})= ik_F \exp\Big(-\frac{i}{2\ell_B^2}\frac{\partial^2}{\partial k_{t} \partial k_n}\Big) S_\varphi(k_n,k_t)
\end{equation}
where we defined
\begin{align}
S_\varphi(k_n,k_t)&=\nonumber\\
 \sum_{\ell=-\infty}^{+\infty} &
\int_{-2/\lambda}^{2/\lambda} dx_n \int_{-2/\Lambda}^{2/\Lambda}dx_t\;  \; G_{\ell,\varphi}(x_n, 0) \; e^{-ik_n x_n-ik_t x_t}
\end{align}
with $\Lambda=2\pi k_F/N$ is a momentum cut-off in the tangential direction to the Fermi surface, with $N \gg1$  a large integer.
Here, we have chosen the phase $\theta_B$ in the Landau gauge, Eq. \eqref{eq:ThetaLandau}.

These integrals can be easily done 
in the limit $\lambda\to 0$ and $\Lambda\to 0$.  However, this limit should be taken carefully. 
Remembering that the number of filled Landau levels is 
$N_L=\lambda \ell_B^2 k_F/2$,  we need  to take $\lambda\to 0$ and $\ell_B\to \infty$ keeping $N_L\gg 1$. Only in this way we  recover the Fermi liquid behaviour  as $B\to 0$. 
The result is, 
\begin{align}
n_{\varphi}(k_n, k_{t})&=k_F\delta(k_t) \Theta(k_F-k_n) 
\sum_{n=-\infty}^{+\infty} \delta_{n,\ell_B^2 k_F(k_n-k_F) }  \nonumber\\
-i& k_F\sin\left\{\frac{1}{\ell_B^2}\frac{\partial^2}{\partial k_{t}\partial k_n}   \right\}
\delta(k_t) \Theta(k_F-k_n) \; .
\label{eq:occupation-varphi}
\end{align}
This equation will be important when computing the algebra of Fermi surface fluctuations in the bosonization  approach. 
Firstly, we observe that $n_\varphi(k_n,k_t)$ is given by very localized distributions. This is a consequence of the thermodynamic limit. Any technical difficulty that could eventually emerge for this reason, can be easily overcome by smoothing  the distributions, keeping the cut-offs $\lambda$ and $\Lambda$ small, but finite. 
An important observation is that $n_\varphi$ is gauge dependent. In particular, the second term in Eq. \eqref{eq:occupation-varphi} comes form the phase $\theta_B$ in the Landau gauge. Then, it should give no contribution in the computation of any observable.

\begin{figure}[ht]
\includegraphics[scale=0.3]{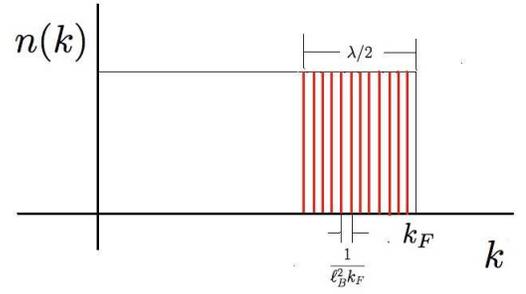}
\caption{Schematic representation of the fermionic occupation number in a weak magnetic field.  The level spacing is $1/\ell_B^2 k_F$, while the number of filled levels near the Fermi surface is $N_L=\lambda\ell_B^2k_F/2$, where $\lambda$ is a momentum cut-off. In the limit of $\ell_B\to\infty$ keeping $\lambda$ fixed, the set of Landau level is dense and we have the usual Fermi gas expression $n(k)=\Theta(k_F-k)$.}
\label{fig:n}
\end{figure}

By replacing Eq. \eqref{eq:occupation-varphi} into Eq. \eqref{eq:n-Int-varphi}  we  can compute the full  occupation number of the fermions. As stated, the second line of Eq. \eqref{eq:occupation-varphi} gives no contribution upon integration over $\varphi$.
When written in terms of the angle $\varphi$, the first line  reads, $n_{\varphi}({\bm k},\varphi)=k_F\delta(k\sin\varphi) \Theta(k_F-k\cos\varphi) 
\sum_n\delta_{n,\ell_B^2 k_F(k\cos\varphi-k_F) }$. Upon integration over $\varphi$ we find, 
\begin{equation}
n({\bm k})= \Theta(-q)\sum_{n=-\infty}^{+\infty} \delta_{n,\ell_B^2 k_F q } + O (\frac{q}{k_F})  \; ,
\end{equation}
where  $q=k-k_F$ with $k=|{\bm k}|$.  The Kronecker deltas  simply indicate the location of the Landau levels,
\begin{equation}
q=-\frac{n}{\ell_B^2 k_F}=-\frac{n}{2} \left(\frac{\omega_c}{\epsilon_F}\right) k_F \; , 
\end{equation} 
which are equally separated in intervals  $\Delta q= 1/\ell_B^2 k_F$. The sum over $n$ is actually limited by the cut-off $\lambda$ to the range  $-N_L\leq n\leq N_L$.
The Heaviside function just  locates the position of the Fermi surface. 
We depict this function in Fig. (\ref{fig:n}). 

In the limit $\ell_B\to \infty$, the  filled  Landau levels,  for $q<\lambda/2$, diverges and become a dense set of points, such that the following limit holds
\begin{equation}
\lim_{\ell_B\to \infty} \sum_{n=-\infty}^{+\infty} \delta_{n,\ell_B^2 k_F(k_F-k) } =\lim_{\ell_B\to \infty} \sum_{\ell=-\infty}^{+\infty} e^{i2\pi\ell\ell_B^2 k_Fq }=1\; , 
\end{equation}
where the last limit is taken in the sense of a distribution, {\em i.\ e.\ }, for huge values of $\ell_B$, terms with $\ell\neq 0$ are strongly oscillating, giving no contribution when applied to smooth test functions. In this regime, the only relevant term is $\ell=0$, 
obtaining the  well known occupation number of a  Fermi gas at zero temperature, 
$
n({\bm k})= \Theta(k_F-k)
$.

An additional consistency check of the asymptotic propagator deduced in this section is  the computation of the density of states   
\begin{equation}
N(\omega)=-\sgn(\omega-\epsilon_F)\frac{1}{\pi} {\rm Im} G_F(\bm{x},\bm{x}, \omega) \; .
\end{equation}
Using  Eq. \eqref{eq:Propagator},  we obtain (see Appendix \ref{Ap:DensityofStates}), 
\begin{equation}
N(\omega-\epsilon_F)=N(0) \sum_n\delta_{n, (\omega-\epsilon_F)/\omega_c} \; , 
\end{equation}
where $N(0)=k_F/v_F$ is the density of states at the Fermi surface of a two-dimensional Fermi gas. 

\section{Bosonization}
\label{Sec:Bosonization}
To proceed with bosonization, we  follow closely the approach outlined in  Refs. [\onlinecite{CastroNeto-1993,CastroNeto-1994,CastroNeto-1995, BaOx2003}]. The main idea is to project the fermion operator into a restricted Hilbert space, built by restricting the momentum space to a small shell around a Fermi surface.  We expect that this restriction correctly capture the low energy and long distance physics at high density, even in the presence of a small magnetic field, provided there is a very large number of Landau levels inside the shell.  

\begin{figure}
\centering
\includegraphics[scale=0.8]{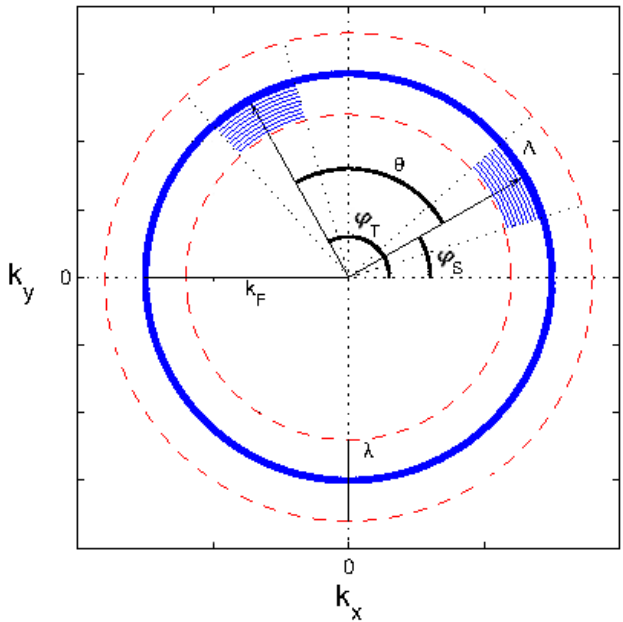}
\caption{Segmentation of a circular Fermi surface of radius $k_F$ in patches of width $\Lambda$ and height $\lambda$. Two patches label by the angles $\varphi_S$ and $\varphi_T$, separated by an angle $\theta$, are displayed. The shaded area represents a set of filled Landau levels inside the patch.}
\label{fig:patches}
\end{figure}

For simplicity, we illustrate de procedure by considering a circular Fermi surface with Fermi momentum $|\bm{k_F}|=k_F$, which we coarse-grain by introducing $N$ patches,  each of width $\Lambda$ and height $\lambda$, as shown in Fig.\ref{fig:patches}.   
The precise shape of each patch is not important. At the end of the calculations, the limit $\Lambda\to 0$, $N\to \infty$
should be taken, with the constraint  that, in the continuum limit,  they add up to the size of the Fermi surface, i.e. $N\Lambda=2\pi k_F$. Each patch is labeled by an integer   $S=1,\ldots,N$ or, equivalently, by an angle 
$\varphi_S= S \Lambda/k_F$,  indicating the angular position of the patch on the Fermi surface with respect to some arbitrary chosen axes (see Fig. \ref{fig:patches}). 
Quite generally, we have that any set of  functions $f_S$, label by $N$ integer numbers,  $S=1,
\ldots, N$, are equivalent, in the continuum limit,  to periodic functions $f(\varphi_S+2\pi)=f(\varphi_S)$, in such a way that
\begin{align}
\frac{\Lambda}{k_F}\sum_S  \; f_S &\to    \int_0^{2\pi} d\varphi_S  f(\varphi_S) 
\nonumber   \\
\frac{k_F}{\Lambda}\left(f_{S+1}-f_S\right) &\to \frac{df(\varphi_S)}{d\varphi_S}  \; .
\end{align} 

We introduce fermionic field operators $\psi_S({\bm x}),\psi^\dagger_S({\bm x})$ on each patch, $S$,  defined as,  
\begin{align}
\psi_S({\bm x})&= \frac{1}{L}e^{-i {\bm k}_S \cdot {\bm x}}\sum_{\bm k}  \Omega_S({\bm k}) \;\; c_{{\bm k}}\; e^{i{\bm k}\cdot {\bm x}} 
\label{eq:psiS-FT}\\
\psi^\dagger_S({\bm x})&=\frac{1}{L}e^{+i{\bm k}_S \cdot {\bm x}}\sum_{\bm k}  \Omega_S({\bm k})\;\;  c^\dagger_{{\bm k}}\; e^{-i{\bm k}\cdot {\bm x}} \; .
\label{eq:psiSdagger-FT}
\end{align}
Here, the operators $c_{\bm k}^\dagger$ ($c_{\bm k}$) create (destroy) a fermion with definite momentum ${\bm k}$ (for the moment, we are ignoring  the spin degree of freedom). Here, ${\bm k}_S$ is the Fermi momentum $\bm{k}_F$, at  the center of the patch $S$, and  $\Omega_S(\bm {k})$ is  a compact support distribution that takes the values: one, for  ${\bm k}$  inside the patch $S$, and zero otherwise. We work in a large but  finite volume, of linear size $L$, and area $L^2$, in such a way that ${\bm k}$ form a dense set of discrete values.  Eventually, we may take the thermodynamic  limit 
$L \to \infty$, for  which $\sum_{\bm k}\to  L^2 \int d^2k$.  We are using the normalization factors in the Fourier transforms, Eqs. \eqref{eq:psiS-FT} and  \eqref{eq:psiSdagger-FT}, in such a way to have the following bare field scaling dimensions:
$[c_{\bm k}]=[c^\dagger_{\bm k}]=1$ and  $[\psi_S({\bm x})]=[\psi^\dagger_S({\bm x})]=L^{-1}$. 

The usual fermionic anticommutation relations 
\begin{equation}
\left\{c^{\dagger}_{{\bm k}},c_{{\bm k}'} \right\}=\delta_{{\bm k},{\bm k}'}
\label{eq:anticommutation-k}
\end{equation}
imply, for the corse-grained fermions, 
\begin{equation}
\left\{\psi_S^{\dagger}({\bm x}),\psi_T({\bm x}')\right\}=\delta_{S,T} \delta^2({\bm x}-{\bm x}') \; .
\label{eq:anticommutation-S}
\end{equation}
The complete Fermi field can be reconstructed by summing over the entire Fermi surface, 
\begin{equation}
\psi({\bm x})=\sum_{S=1}^N  e^{-i{\bm k}_S\cdot {\bm x}} \psi_S({\bm x})\; .
\label{eq:CompleteFermion}
\end{equation}
The patch fermions $\psi_S({\bm x})$ belong to a mixed coordinate-momentum representation,  where ${\bm x}$ is a configurational space variable and $S$, or equivalently,  $\varphi_S$ is a momentum variable.  
Eq. \eqref{eq:CompleteFermion}, is the generalization of the  well known one dimensional expression, 
$
\psi(x)=e^{i k_F x} \psi_R(x)+e^{-i k_F x} \psi_L(x)
$, that relates the microscopic field operator to the slowly varying right and left moving fields. Clearly, $\psi_S(x)$ generalizes the concept of left, $\psi_L$, and right, $\psi_R$, moving  fermions. To be consistent with quantum mechanics, i.e. with the Heisenberg uncertainty principle, the fields $\psi_S({\bm x})$ are slowly varying functions on a length scale $\sim k^{-1}_F$ (the mean distance between fermions in real space).  

The low-energy effective Hamiltonian  can be obtained  by projecting the free fermion Hamiltonian of  Eq. \eqref{eq:Hamiltonian} to the restricted Hilbert space spanned by states in the small shell around the Fermi surface. We find, 
\begin{equation}
H_F=i v_F\sum_S\int d^2x\;  \psi^\dagger_S({\bm x})\;\left( \hat {\bm n}_S\cdot {\bm D}\right) \psi_S({\bm x}) \; ,
\label{eq:Fermionic-Hamiltonian-S0}
\end{equation} 
where $v_F$ is the Fermi velocity, $\hat {\bm n}_S$ is  a constant unit vector normal to the Fermi surface, pointing outward in the patch $S$, and we have represented the covariant derivative acting on the  projected space spanned by $\psi_S({\bm x})\equiv\psi({\bm k}_S,{\bm x})$, as   
\begin{equation}
D_i=\nabla_i +i  k_{S,i}-\frac{1}{\ell_B^2} \epsilon_{ij} \frac{\partial}{\partial k_{S,j}} \; .
\label{eq:CDS}
\end{equation}
Here, the covariant derivative has two parts. The first term acts on the slowly varying position variable ${\bm x}$, while  the second part is in the momentum representation and acts on the Fermi surface. We will show that, since both parts commute with each other, this scale separation has important consequences on the symmetries of the system.    

Eq. \eqref{eq:CDS} was obtained by fixing the symmetric gauge   $A_i=\frac{B}{2}\epsilon_{ij} x_j$. The term proportional to $i\bm{k}_S$ contributes to the energy of the ground state ($\epsilon_F=v_F k_F$) and can be ignored, since it cancels out upon normal ordering the Hamiltonian with respect to the filled Fermi sea. We decided to keep it in the definition of the covariant derivative, since it is important to study symmetries of the low energy Hamiltonian, as we will see in the next section.  

To obtain the effective low energy Hamiltonian of Eq. \eqref{eq:Fermionic-Hamiltonian-S0}, we  linearized the dispersion relation in each patch, considering that the momentum excitations have $|{\bm k}-{\bm k}_F|<\lambda<\Lambda\ll  k_F$. We have also ignored terms of the order $1/(\ell_B k_F)^2\ll 1$. In other words, we have neglected sub-leading terms in $\omega_c/v_Fk_F\ll 1$, meaning that the energy difference between Landau levels is much smaller than the Fermi energy. The Hamiltonian $H_F$ is the generalization of the usual low-energy Hamiltonian considered in bosonization of Fermi liquids.\cite{CastroNeto-1993,CastroNeto-1994,CastroNeto-1995,houghton-1993,houghton-1994,houghton-2000,BaOx2003}  In the presence of a weak magnetic field the chiral fermionic components are coupled with the derivative term proportional to $1/(\ell_B^2 k_F)$.   

Particle-hole excitations are created  by acting with the operator   $\hat n_{\bm k}({\bm q})=c^\dagger_{{\bm k}+{\bm q}/2}c_{{\bm k}-{\bm q}/2}$ on the reference state $|FS\rangle$.
 Deformations of the Fermi surface can be parametrized by a normal ordered particle-hole operator, smeared  at each patch. We define 
\begin{align}
\delta n_S({\bm q})&=\sum_k \Omega_S({\bm k}-{\bm q}/2)\Omega_S({\bm k}+{\bm q}/2)
\label{eq:delta-nS}\\
&\times \left\{ c^\dagger_{{\bm k}+{\bm q}/2}c_{{\bm k}-{\bm q}/2}-\langle FS| c^\dagger_{{\bm k}+{\bm q}/2}c_{{\bm k}-{\bm q}/2}|FS\rangle   \right\}\; .
\nonumber
\end{align}
As before, $\Omega_S({\bm k})$ is a distribution with support inside the patch $S$. By definition, in the ground state, with an undistorted Fermi surface,  $\langle FS|\delta n_S|FS\rangle=0$, since the operator is normal ordered with respect to this state. 

The dynamics of the Fermi surface  is governed by the Heisenberg equation
\begin{equation}
\frac{\partial \delta n_S(\bm{x})}{\partial t}=i \left[ H_F,\delta n_S(\bm{x})\right] \; .
\end{equation}
Using the fermionic anticommutator algebra, Eq. \eqref{eq:anticommutation-k}, we find that this equation of motion has the form of a collisionless Boltzmann type equation  
\begin{equation}
\frac{\partial \delta n_S(\bm{x})}{\partial t}+{\bm v}_S\cdot {\bm \nabla}\delta n_S(\bm{x})-\omega_c \frac{\partial \delta n_S(\bm{x})}{\partial \varphi_S}=0 \; .
\label{eq:landau-Silin}
\end{equation}
While the second term rules the excitations normal to the Fermi surface, the last term, coming form the Lorentz force, induces an evolution of the patch variable around the Fermi surface.  This equation is a particular case of the Landau-Silin\cite{nozieres-1999} equation obtained by means of a phenomenological treatment of a Fermi liquid in an electromagnetic field.  
A similar quantum Boltzmann equation was  deduced previously in the context of the quantum Hall effect at filling factor $\nu\sim 1/2$, by  applying  non-equilibrium Green function techniques to a fermion model coupled to a Chern-Simons field.\cite{Kim-1995}   
More recently, a related magneto transport equation with similar structure was deduced in a context of disorder strange metals.\cite{PaMcGreeArSa-2018}

The question is whether we can get this equation of motion from a pure Bosonic approach. To answer it, we firstly compute the commutation relations of $\delta n_S(\bm{x})$.  Using Eq. \eqref{eq:delta-nS} and \eqref{eq:anticommutation-k} we find, 
\begin{align} 
&[\delta n_S({\bm q}), \delta n_T(-{\bm q}')]= \\
&\delta_{{\bm q},{\bm q}'}
  \sum_{\bm k} \Omega_S({\bm k}-{\bm q}/2)\Omega_T({\bm k}+{\bm q}'/2)
\left(
n_{\varphi_S}({\bm q})-n_{\varphi_T}(-{\bm q}')\right)
\nonumber
\end{align}
where $n_{\varphi_S}({\bm q})$ and $n_{\varphi_T}(-{\bm q})$ are the occupation numbers at the  patches $S$ and $T$ respectively, computed in the previous section (Eq. (\ref{eq:occupation-varphi})). Expanding these quantities for small values of ${\bm q}$, we find,
\begin{equation}
[\delta n_{S}({\bm q}), \delta n_{T}(-{\bm q}')]=\left(\frac{L^2\Lambda}{2\pi}\right)
\tilde D_S({\bm q})\delta_{S,T}  \delta_{{\bm q},{\bm q}'}
\label{eq:Schwinger-algebra-q}
\end{equation}
where $\tilde D_S({\bm q})$ is given by 
\begin{equation}
\tilde D_S({\bm q})= \hat {\bm n}_S\cdot{\bm q}-\frac{i}{\ell_B^2k_F}\frac{\partial}{\partial \varphi_S} 
\label{eq:CovariantDerivative-q}
\end{equation}
Fourier transforming  Eq. (\ref{eq:Schwinger-algebra-q}), we have the commutation relation in real space,  
\begin{equation}
[\delta n_{S}({\bm x}), \delta n_{T}({\bm y})]= -\left(\frac{\Lambda}{2\pi}\right)
  i D_S\left[
\delta^2({\bm x}-{\bm y})\delta_{S,T}\right] \; ,
\label{eq:Schwinger-algebra}
\end{equation}
where 
\begin{equation}
D_S=\hat {\bm n}_S\cdot {\bm \nabla}+ \frac{1}{\ell_B^2 k_F}\frac{\partial}{\partial \varphi_S} \; .
\label{eq:CovariantDerivative} 
\end{equation}
This equation is the key of the bosonization procedure, and, for this reason, it is important  to comment on its meaning and its range of applicability. 

In Eq.\eqref{eq:CovariantDerivative} 
$D_S$ is essentially the normal component of the covariant derivative at each patch $S$. 
At zero magnetic field, the expression $D_S= \hat {\bm n}_S\cdot {\bm \nabla}$ reduces to the usual Schwinger term analogous to the one  found  in one-dimensional bosonization (where $S= R, L$) and in bosonization of Fermi liquids.\cite{CastroNeto-1993,houghton-1993,haldane-1994} The effect of the magnetic field is to produce a new  term tangent to the Fermi surface. In this way, in the presence of a magnetic field,  the commutation relations are no longer diagonal in the patch basis, and hence mix states on different patches of the Fermi surface. The algebra of Eq. \eqref{eq:Schwinger-algebra} can be regarded as a {\em covariant Schwinger algebra}. Essentially the same commutation relations have been recently proposed, computed by means of semiclassical arguments, in the (quite different) context of fractional Hall states.\cite{Golkar-2016}  

As usual in the treatment of multidimensional bosonization, in order to close the algebra  we have discarded terms proportional to the ratio $\lambda/\Lambda \ll 1$. Then, the limit of  a large number of patches  $N$ should be taken satisfying  $\lambda<\Lambda\ll k_F$. While this is well stablished in the bosonization of a Fermi liquid, special care should be taken in the presence of a magnetic field. Indeed, we need to guarantee that, even in the limit of very small $\lambda$, there should be  a huge number of Landau levels $N_L=\lambda\ell_B^2k_F/2\gg 1$ inside the momentum shell around $k_F$.  This requirement thus relates the number of patches $N$ with the number of Landau levels $N_L$. In fact, the limits $N\to \infty, \ell_B\to \infty$ do not commute. 
To clarify this point, suppose that we may (wrongly)  take the limit $N\to \infty$, fixing the magnetic length $\ell_B$. Then,  the width of the energy shell  around the Fermi energy gets smaller than the gap between  Landau levels, $\Delta \epsilon <\omega_c$. In this case, the concept of a Fermi liquid with a well defined Fermi surface is lost. As a last interesting remark, notice that in Eq. \eqref{eq:Schwinger-algebra}, the magnetic field contribution is exact, {\em i.\  e.\  }, there are no  higher order correction terms in a $1/\ell_Bk_F$ expansion.  

Having stablished the bosonic character of the field $\delta n_S$, we can introduce the local quadratic Hamiltonian,
\begin{equation}
H_B= \frac{\pi v_F}{L^2\Lambda}\sum_S \int d^2x\;   :\delta n_S^2(\bm{x}): \; ,
\label{eq:Bosonic-Hamiltonian}
\end{equation}
where the colons mean normal order with respect to the filled Fermi sea state, $|FS\rangle$.
By computing the Heisenberg equation of motion, 
\begin{equation}
\frac{\partial \delta n_S(\bm{x})}{\partial t}=i \left[ H_B,\delta n_S(\bm{x})\right] \; ,
\end{equation}
using the bosonic algebra of Eq. \eqref{eq:Schwinger-algebra}, we  find that it is exactly the same equation, Eq. \eqref{eq:landau-Silin}, that we have  found above  by direct computation in  terms of the fermionic fields.

Thus, we  stablished that, the low energy fermionic Hamiltonian  $H_F$, of Eq. \eqref{eq:Fermionic-Hamiltonian-S0},  with the standard anti-commutation relations, leads to the same equation of motion for the particle-hole excitations that the one derived by using  the quadratic bosonic Hamiltonian $H_B$, of Eq. \eqref{eq:Bosonic-Hamiltonian}, with the bosonic {\em covariant Schwinger algebra}, Eq. \eqref{eq:Schwinger-algebra}. This is the main result of this section. 

It is convenient to have a representation of the Fermi surface deformations, and the low energy Hamiltonian, in terms of canonical Bose fields.  
Since $\delta n_S(\bm{q})$ does not annihilate the ground state, we define the following operators:
\begin{align}
a_S({\bm q})&=\sqrt{\frac{2\pi}{L^2\Lambda}}
\label{eq:a}  \\ 
\times  \sum_{S'}&   \Big[\alpha_{S,S'}({\bm q}) \delta n_{S'}({\bm q})\Theta(q_{n'})+\alpha_{S,S'}^\dagger({\bm q}) \delta n_{S'}(-{\bm q})\Theta(-q_{n'})\Big] \nonumber \\
a^\dagger_S({\bm q})&=\sqrt{\frac{2\pi}{L^2\Lambda}}  
\label{eq:adagger} \\
\times \sum_{S'} &  \Big[\alpha^{\dagger}_{S,S'}({\bm q}) \delta n_{S'}(-{\bm q})\Theta(q_{n'})+\alpha_{S,S'}({\bm q})\delta n_{S'}({\bm q})\Theta(-q_{n'})\Big] \nonumber 
\end{align}
where $\alpha_{S,S'}({\bm q})$ is a set of matrices  to be determined, $q_{n'}=\hat {\bm n}_{S'}\cdot {\bm q}$, and $\Theta$ is the usual Heaviside distribution. It is immediate to verify that
\begin{equation}
a_S({\bm q})\left| FS\rangle \right.=0\; .
\label{eq:KillFS}
\end{equation}
In addition, the requirement that the bosonic operators $a_S$ and $a^\dagger_S$ satisfy canonical commutation relations
\begin{equation}
[a_S({\bm q}),a^\dagger_T({\bm q}')]=\delta_{S,T}\left(\delta_{{\bm q},{\bm q}'}+\delta_{{\bm q},-{\bm q}'}\right)  \; , 
\label{eq:canonical}
\end{equation}
fixes the matrix $\alpha_{S,T}({\bm q})$ to satisfy, 
\begin{equation}
\sum_{T'} \alpha^\dagger_{T,T'}({\bm q})\alpha_{T',S}({\bm q})=\tilde D^{-1}_{S,T}({\bm q}), 
\label{eq:alpha}
\end{equation}
where  $\tilde D^{-1}_{S,S'}({\bm q})$ is the Green function 
\begin{equation}
\left\{\hat {\bm n}_S\cdot{\bm q}-\frac{i}{\ell_B^2k_F}\frac{\partial}{\partial \varphi_S}\right\}\tilde D^{-1}_{S,S'}({\bm q})=\delta_{S,S'}\; .
\label{eq:GreenCovariant}
\end{equation}
We will see that this Green  function plays a central role of the bosonization theory. It will recurrently appear in the structure of the bosonization construction, as well as in the computation of all correlation functions. In Appendix \ref{Ap:GFD}, we study in detail its mathematical properties.

Inverting Eqs. \eqref{eq:a} and \eqref{eq:adagger}, we obtain the desired relation that describes the Fermi surface deformation in terms of a set of canonical harmonic oscillators, 
\begin{align}
&\delta n_{S}({\bm q})=\frac{L}{2\pi}\Lambda^{1/2}\;\; 
\label{eq:deltan-a} \\
&  \times \sum_{S'}\left[\beta_{S,S'}a_{S'}({\bm q})\Theta(q_{n'})+\beta_{S,S'}^\dagger a_{S'}^\dagger({\bm q})\Theta(-q_{n'})\right] \nonumber  
\end{align}
where we introduced the matrix $\beta=\alpha^{-1}$. We see that, a local deformation of the Fermi surface $\delta n_S({\bm q})$ is given by a linear superposition of canonical harmonic oscillators  around the Fermi surface. 

Replacing Eq. \eqref{eq:deltan-a} into Eq. \eqref{eq:Bosonic-Hamiltonian}, we obtain the bosonic Hamiltonian in terms of canonical harmonic oscillator fields,   
\begin{equation}
H_B=v_F \sum_{S}\sum_{{\bm q}}\;  a^\dagger_S({\bm q})\;  \tilde D_S({\bm q})\;  a_S({\bm q})  \;.
\label{eq:BHa}
\end{equation}
As expected, the quadratic bosonic Hamiltonian is not diagonal in the patch basis, when written in terms of canonical fields.

\section{Symmetries of the low-energy effective Hamiltonian} 
\label{Sec:Symmeties}

Symmetries of a microscopic model should be checked  in an effective low-energy model. On the other hand,  low energy Hamiltonians often  tend to exhibit larger  symmetries,  that are broken at a more microscopic level.  A standard example is rotational symmetry which is broken down to the (discrete) point group symmetry of a lattice but  becomes continuous at a  critical point. Another  example more relevant here is the case of a Fermi liquid. A microscopic interacting fermionic Hamiltonian should have global gauge $U(1)$ invariance, related with the conservation of particle number. Of course, any low energy effective theory, such as Landau theory of Fermi liquids, or higher dimensional bosonization, should at least have this same symmetry (provided the total particle number is still conserved.) 

However, new symmetries can emerge at low energies.  
In fact, in the Landau theory of the Fermi liquid the low energy Hamiltonian is invariant under a global phase transformation {\em at each point (patch) of the Fermi surface}. Then, the system with $N$ patches has a $U(1)^N$ symmetry, related with the particle number conservation at each patch.  In the thermodynamic limit, the theory formally has a huge $U(1)^\infty$ invariance. In the language of the renormalization group, we say that the microscopic interactions that break $U(1)^\infty$ symmetry are irrelevant operators at the Fermi liquid fixed point.\cite{CastroNeto-1995,Shankar1994,Polchinski-1992}   We will now see that something similar occurs in the Fermi gas in a magnetic field. 

\subsection{Global U(1) symmetry}

Let us begin with the symmetries of the projected low energy Hamiltonian.
Since the particle number is globally conserved, the Hamiltonian should be invariant under global phase transformations, 
$\psi'_S(\bm{x})=\exp(i\alpha) \psi_S(\bm{x})$, for $S=1,\ldots, N$ and $\alpha$  a constant.
However, unlike the Fermi liquid fixed point which has a $U(1)$ symmetry {\em for each patch}, the Hamiltonian of a Fermi gas coupled to an external uniform magnetic field  is not invariant under phase transformation on each patch, $\psi_S(\bm{x})\to\exp(i\alpha_S) \psi_S(\bm{x})$, since the magnetic field breaks this symmetry, thus implying that the charge is not conserved separately at each patch.

This effect has a close relation with the chiral anomaly. If we take for instance a one-dimensional sub-system by considering two opposite patches, say $S_1$ and $-S_1$, the Hamiltonian is invariant under  the chiral transformation $\psi'_{\pm S_1}(\bm{x})=\exp(\pm i\alpha) \psi_{\pm S_1}(\bm{x})$. In the presence of an external electric field along ${\bm n}_{S_1}$ the divergence of the chiral current is not zero, meaning that the charge is not conserved in each patch independently.  In this case, the system develops a charge density way, breaking translation invariance\cite{Fradkin-Book}. This chiral anomaly is a direct consequence of the Schwinger term in the $U(1)$ current algebra. The magnetic field has no effect on each one-dimensional sub-system. However, it breaks translation symmetry to  magnetic translations. To take this effect into account, the chiral anomaly (the Schwinger term) should be modified by the magnetic field, mixing different patches as we have already described in Eq. (\ref{eq:Schwinger-algebra}). This term should transfer charge from path to path along the Fermi surface.

At the quantum level, the generator of the global $U(1)$ symmetry  is the particle number  operator $\Delta N$, normal ordered with the filled Fermi sea reference state $|FG\rangle$.  In the bosonic representation, it  can be written as
\begin{equation}
\Delta N=\sum_S\Delta N_S=\sum_S\int d^2x\; \delta n_S(\bm{x}) \; ,
\label{eq:NS}
\end{equation} 
where we have introduced the number operator at each patch $\Delta N_S$.  
Using the bosonized Hamiltonian $H_B$,  given by Eq. \eqref{eq:Bosonic-Hamiltonian},  and the covariant Schwinger algebra of Eq. \eqref{eq:Schwinger-algebra},  it is simple to derive the commutation relation 
$ 
[\Delta N_S, H_B]= -i\omega_c \frac{d \Delta N_S}{d\varphi_S}
$. 
Consequently, the equation of motion for the particle number at each patch is
\begin{equation}
\frac{\partial \Delta N_S}{\partial t}+\omega_c \frac{\partial \Delta N_S}{\partial \varphi_S}=0\;.
\end{equation}
A general solution,  $\Delta N_S\equiv\Delta N_S(\varphi_S-\omega_c t)$, is a local charge perturbation wrapping around the Fermi surface with velocity $\omega_c/k_F$.

In the absence of magnetic field, $\omega_c\to 0$,  $[\Delta N_S, H_B]=0$, and the system is $U(1)$ invariant at each patch, independently. However, even for a weak  magnetic field, this huge symmetry is broken. Of course, by summing up over the patches of the hole Fermi surface, we obtain  $[\Delta N, H_B]=0$, as it should be, since the total charge is conserved. 

\subsection{Magnetic translations}

The full Hamiltonian, Eq. \eqref{eq:Hamiltonian}, is not invariant under translations, due to the presence of the uniform magnetic field. Instead, it is invariant under the transformation 
$\psi(\bm{x})\to \exp\{-i (B/2){\bm a} \times {\bm x}\}\psi({\bm x}+{\bm a})$, that characterizes the {\em group of magnetic translations}. Its infinitesimal generators are $D^*_i=\partial_i-i e A_i$, which satisfy
\begin{align}
[ D^*_i, D^*_j ] &=i B \epsilon_{ij} \; , 
 \label{eq:D*D*B} \\
[ D^*_i, D_j ] &= 0  \; .
\label{eq:D*D0}
\end{align}   
Elements  of the group, {\em i.\ e.\ },  finite magnetic translation labeled by a vector ${\bm a}$,  $T_{{\bm a}}=\exp({\bm a}\cdot {\bm D}^*)$,  satisfy the algebra
\begin{equation}
[T_{{\bm a}}, T_{{\bm b}}]=2i\sin\left(\frac{B}{2} {\hat {\bm z}}\cdot  \left({\bm a}\times {\bm b} \right)\right) T_{{\bm a}+{\bm b}}\, .
\end{equation}
Using Eqs. \eqref{eq:Hamiltonian}, \eqref{eq:D*D*B} and \eqref{eq:D*D0}, we easily show that  $[T_{{\bm a}}, H]=0$, confirming that this is indeed a symmetry. Evidently, this symmetry should also be obeyed by the (patched) low-energy Hamiltonian. 

It is instructive to see how the infinitesimal generators of magnetic translations are represented on the fermionic basis for each patch, $\psi_S(\bm{x})$. 
Considering the projection of $D^*_i$ on the small shell around the Fermi surface, we have 
\begin{equation}
 D^*_i= \nabla_i+i  k_{S,i}+\frac{1}{\ell_B^2} \epsilon_{ij} \frac{\partial}{\partial k_{S,j}}\; .
\end{equation}
It is very easy to check that these operators in fact satisfy the algebra of Eqs. \eqref{eq:D*D*B} and \eqref{eq:D*D0} . In fact, Eq. \eqref{eq:D*D0} guarantees that these generators commute with the Hamiltonian, $[D^*_i, H_F]=0$, implying that the low energy Hamiltonian is invariant under the infinitesimal transformation 
\begin{equation}
\delta \psi_S({\bm x})=  \left({\bm a}\cdot {\bm \nabla}+i{\bm a}\cdot {\bm k}_{S}+\frac{1}{\ell_B^2}   {\bm a} \times \frac{\partial}{\partial {\bm k}_{S}}\right)\psi_S({\bm x}) \; .
\label{eq:MTS-inf}
\end{equation} 
The action of  a finite magnetic translation on a patch fermion is
\begin{equation}
\psi'_S({\bm x})=T_{{\bm a}}\psi_S({\bm x})=e^{i {\bm k}_S\cdot {\bm a}} \psi_{S+\Delta S}({\bm x}+{\bm a})
\label{eq:MTS}
\end{equation} 
where, $\Delta S=  \hat {\bm n}_S\cdot {\bm a}/\ell_B^2 k_F$. The net effect on $\psi_S$ is, not only to translate the position ${\bm  x}\to {\bm x}+{\bm a}$, but also to shift the patch parameter $S$ by $\Delta S$. The transformation also changes the phase by an amount ${\bm k}_S\cdot {\bm a}$.

To  clarify that the transformation of Eq. \eqref{eq:MTS} is in fact a magnetic translation, it is necessary to  reconstruct the full  fermion operator by adding up the contributions over the patches of the whole Fermi surface. We then have,  
\begin{align}
\psi'({\bm x})
&=
\sum_S  e^{i {\bm k}_S \cdot {\bm x} } e^{i {\bm k}_S\cdot {\bm a}} \psi_{S+\Delta S}({\bm x}+{\bm a})
\nonumber \\
&=\sum_S e^{i\bm{ k}_{S-\Delta S}\cdot( \bm{x}+\bm{a})} \psi_S(\bm{x}+\bm{a}) \; ,
\end{align}
where, in the last equality,  we have shifted $S\to S-\Delta S$.
In the limit of small magnetic fields, $1/\ell_B k_F\ll 1$,  we can write ${\bm k}_{S-\Delta S}\cdot {\bm x}=
{\bm k}_S \cdot {\bm x}-({\bm x}\times {\bm a})\cdot {\hat {\bm z}} B/2$, and then   
\begin{equation}
\psi'({\bm x})=e^{i \frac{B}{2}{\hat {\bm z}} \cdot ({\bm a}\times {\bm x})} \psi({\bm x}+{\bm a}),
\end{equation}
which is precisely a magnetic translation of the full fermion operator.

Our next task is construct a bosonic representation of  the  magnetic translation generators. 
It is known that the elements of the group can be represented as non-trivial linear superposition of fermion bilinears.\cite{Shankar2012,Chamon2012} In our case, it is more convenient to use a non-linear representation in terms of the bosonic excitations $\delta n_S$ or, equivalently, $a_S, a^\dagger_S$. The desired representation of the generators of magnetic translations  is
\begin{equation}
{\bm \Pi}
=\sqrt{\frac{\Lambda}{k_F}}\frac{1}{k_FL}\sum_S \int d^2x \left( i {\bm k}_S \delta n_S({\bm x})+\frac{1}{2}\hat {\bm n}_S :\!\delta n^2_S({\bm x})\!:\right)
\label{eq:DdeltanS}
\end{equation}
which coincides with the canonical momentum of a Fermi liquid.\cite{haldane-1994} The covariant Schwinger algebra, Eq. \eqref{eq:Schwinger-algebra}, guarantees that this expression satisfies the magnetic translation algebra, Eq. \eqref{eq:D*D*B}.
It is possible to rewrite Eq. \eqref{eq:DdeltanS} in terms of the canonical bosonic fields $a_S$ and $a^\dagger_S$. However, in that representation, ${\bm \Pi}$ is not diagonal in the patch variable $S$.  

\subsection{Translations}

As anticipated, the low-energy Hamiltonian, $H_F$,  has more symmetries than the full Hamiltonian $H$. In fact, $H_F$ is also invariant under  translations $\psi_S'({\bm x})=\psi_S({\bm x}+{\bm a})$. This happens because the projection process brings about a splitting of scales, as was already observed in the structure of the covariant derivative, Eq. \eqref{eq:CDS}.  Indeed,  the rapidly oscillating part, proportional to ${\bm k}_F\cdot {\bm x}$, has been extracted from  the definition of $\psi_S({\bm x})$, leading to a slowly varying field on scales $|{\bm x}|\gg k_F^{-1}$.  The infinitesimal generator of translations, ${\bm P}=-i{\bm \nabla}$, commutes with all the other symmetry generators and with the Hamiltonian,  
{\em i.\ e.\ }, $[P_i, P_j]=[P_i, D^*_j]=[P_i, D_j]=[P_i, H_F]=0$.   The bosonic representation of ${\bm P}$ is written in a simpler way in terms of   the bosonic fields $a_S$ and $a_S^\dagger$,
\begin{equation}
{\bm P}=\sum_S\int d^2q \; {\bm q}\; a^\dagger_S({\bm q}) a_S({\bm q})  \;.
\label{eq:PS}
\end{equation}
Notice that, if rewritten in terms of the fields $\delta n_S$, this is a non-local function of the patch variable $S$.

\section{The Fermion Operator} 
\label{Sec:FermionOperator}

To push forward the bosonization program we need to build the fermion  operator in terms of canonical bosonic fields. 
The fermion operator should satisfy the following commutation relations, 
\begin{align}
[\Delta N, \psi_S({\bm x})]&=-\psi_S({\bm x})  
\nonumber \\
[\Pi_i, \psi_S({\bm x})]&=\left(\nabla_i+i  k_{S,i}+\frac{1}{\ell_B^2} \epsilon_{ij} \frac{\partial}{\partial k_{S,j}}\right)\!\psi_S({\bm x})  
\nonumber \\
[P_i, \psi_S({\bm x})]&=-i \nabla_i \psi_S({\bm x})  \;,
\end{align}
where $\Delta N$, ${\bm \Pi}$ and ${\bm P}$ are the infinitesimal generators of the main symmetries of the low energy system, already described in the previous section. Their bosonic representation is given by Eqs. \eqref{eq:NS}, \eqref{eq:DdeltanS} and \eqref{eq:PS}, respectively. 

Interestingly, these three commutation relations are simultaneously satisfied by simply using,  
\begin{equation}
[\psi_S({\bm x}), \delta n_T({\bm y}) ]=\delta({\bm x}-{\bm y})\delta_{S,T} \psi_S({\bm x})  \; .
\label{eq:psideltan}
\end{equation}
For consistency,  Eq. (\ref{eq:psideltan}) can be independently confirmed by means of the definition of $\delta n_S$  and the usual fermionic algebra.

Similarly to the Fermi liquid case,\cite{CastroNeto-1995} we propose the following ansatz for the  fermion operator,   
\begin{equation}
\psi_S({\bm x})= \frac{1}{2\pi}\sqrt{\Lambda\lambda}\;  K_S\;  e^{i\phi_S({\bm x})} \; ,
\label{eq:Fermion-Operator}
\end{equation}
where $K_S$ are the Klein factors (that insure anticommutation relations of the fermion operators at different patches), and $\phi_S({\bm x})$ is a bosonic phase field. Both operators   will be determined below.

Upon replacing Eq. \eqref{eq:Fermion-Operator} into Eq.\eqref{eq:psideltan}, and additionally assuming that $[K_S, \delta n_T]=0$, and $[\phi_S({\bm x}), \delta n_T({\bm x}')]= c$, where $c$ is a complex number,  we find that the bosonic field $\phi_S({\bm x})$ must obey
\begin{equation}
[\delta n_S({\bm y}),\phi_T({\bm x})]= -i \delta({\bm x}-{\bm y})  \delta_{S,T} \; .
\label{eq:nSphi}
\end{equation} 
Evidently, the fermionic phase, $\phi_S({\bm x})$ is the canonical conjugate field to $\delta n_S({\bm x})$.  
This algebra is then exactly solved by defining 
\begin{equation}
\delta n_S({\bm x})=\frac{\Lambda}{(2\pi)^2} D_S \phi_S({\bm x}) \; , 
\label{eq:deltanphi}
\end{equation}
with the commutation relations, 
\begin{equation}
[\phi_S({\bm x}),\phi_T({\bm y})]=\frac{(2\pi)^2}{\Lambda} D^{-1}_{S,T}({\bm x}-{\bm y}) \; .
\label{eq:phicr}
\end{equation}

Eqs. \eqref{eq:deltanphi} and \eqref{eq:phicr} are an elegant (and important) generalization of the usual expressions of bosonization of  Fermi liquids. Indeed, in the limit $\omega_c\to 0$,  the covariant derivative becomes $D_S\to \hat {\bm n}_S\cdot {\bm \nabla}$ and
the Green function approaches the limit $D^{-1}_{S,T}({\bm x}-{\bm y})\to \sgn({\bm x}-{\bm y})\delta_{S,T}$, thus recovering previous existing  results.\cite{CastroNeto-1993,CastroNeto-1994,CastroNeto-1995,houghton-1993,houghton-1994} 

By inverting Eq. \eqref{eq:deltanphi} we  obtain the desired relation between the  phase $\phi_S({\bm x})$ of the fermionic operator and the bosonic particle-hole excitations, 
\begin{equation} 
\phi_S({\bm x})=-i\left(\frac{2\pi}{L}\right)^2\frac{1}{\Lambda}\sum_{\bm q} e^{-i{\bm q}\cdot {\bm x}}\sum_{S'} \tilde D^{-1}_{S,S'}({\bm q}) \delta n_{S'}(-{\bm q}) \;.
\label{eq:phideltan}
\end{equation}
We can now see the main change between bosonization in the absence of a magnetic field and with the application of a weak enough field.  In the former case, the phase of the fermionic patch operator is a local function of the density fluctuation $\delta n_S$. In the latter, instead, the phase at a single patch $S$ is defined by the contribution of $\delta n_S$ on the entire Fermi surface. The structure of the patch superposition is coded in the  Green function of the normal covariant derivative, $\tilde D^{-1}_{S,S'}$.

It is useful to rewrite the   phase of the fermionic field in terms of canonical fields operators. 
By replacing Eq. \eqref{eq:deltan-a} into Eq. \eqref{eq:phideltan}, we  obtain 
\begin{align}
\phi_S(\bm{x})&=-i\frac{2\pi}{L\Lambda^{1/2}} \sum_{\bm{q}} e^{-i \bm{q}\cdot\bm{x}}    
\label{eq:phi-a} \\
\times \sum_{S'}
\Big\{ &
\alpha_{S,S'}(\bm{q}) a^\dagger_{S'}(\bm{q})\Theta(\bm{q}_{n'})-\alpha^\dagger_{S,S'}(-\bm{q})a_{S'}(\bm{q})\Theta(-\bm{q}_{n'})
\Big\} \; .
\nonumber
\end{align}
Here, the phase operator is written in terms of a coherent superposition of harmonic oscillators,  defined on the whole  Fermi surface. The corresponding weight  is given by the functions $\alpha_{S,S'}$,  that can be roughly thought as the square root of the  Green function, $\alpha\sim \tilde D_S^{-1/2}$. The precise definition of $\alpha_{S,S'}$ is given in Eq. \eqref{eq:alpha}. In the limit of zero magnetic field, the Green function becomes local in the patch basis, and Eq. \eqref{eq:phi-a} reduces to the usual local expression.\cite{CastroNeto-1995,houghton-2000,BaOx2003}    

The Klein factors $K_S$ should be built in order to insure anticommutation relations between fermions defined on  different patches.  Consider for instance, 
\begin{equation}
K_S= \lim_{{\bm q} \to 0}e^{i \pi \sum_{S'=1}^{S-1} \delta n_{S'}({\bm q})}
\end{equation}
where we have chosen an arbitrary reference $S=1$ and we have  ordered  the patches counterclockwise. 
It is then straightforward to show that the Klein factors commute with each other,  
$[K_S,K_T]=0$, and that they are their own inverse, i.e.
$K_S K_S^\dagger=1$.
Moreover,  by using Eq. \eqref{eq:nSphi}, we have, 
\begin{equation}
K_T^\dagger \psi_S K_T =
\begin{cases}
-\psi_S  &  \textrm{~~if~~}    T> S  \\
+\psi_S  &  \textrm{~~if~~}   T\le  S   
\end{cases}
\end{equation}
This construction completes the definition of the fermion operator.

\subsection{The  fermion propagator at equal times}

As an example, and to check the consistency of the construction, let us compute the fermionic equal time expectation value 
on the same patch.
Using the definition of the fermion operator, Eq. \eqref{eq:Fermion-Operator}, we have
\begin{equation}
\langle \psi^\dagger_S({\bm x})\psi_S(0)\rangle=\frac{\lambda\Lambda}{(2\pi)^2}\langle e^{-i\phi_S({\bm x})}e^{i\phi_S(0)}\rangle  \; .
\label{eq:eqtime-patch-fermion}
\end{equation}
Using the Baker-Hausdorff formula, 
\begin{equation}
e^{\hat A}e^{\hat B}=: e^{\hat A+\hat B}: e^{\langle\hat A\hat B+(1/2)(\hat A^2+\hat B^2)\rangle }
\end{equation}
we can write, 
\begin{equation}
\langle \psi^\dagger_S({\bm x})\psi_S(0)\rangle=
\frac{\lambda\Lambda}{(2\pi)^2} e^{G_{\phi}(S, {\bm x})} \; 
\label{eq:equaltime}
\end{equation}
where we used the notation
\begin{align}
G_{\phi}(S,{\bm x})&=\langle FS| \phi_S({\bm x})\phi_S(0)-\phi_S^2(0)|FS\rangle \nonumber\\
&=-\frac{1}{2} \langle FS| (\phi_S({\bm x})-\phi_S(0))^2|FS\rangle \; .
\end{align}
Using the representation of the bosonic field $\phi_S$ in terms of harmonic oscillators, Eqs. \eqref{eq:phi-a}, and the fact that $a_S|FS\rangle=0$, it is simple to obtain,  
\begin{equation}
G_{\phi}(S,{\bm x})=\left(\frac{2\pi}{L}\right)^2\frac{1}{\Lambda}\sum_{{\bm q},q_n>0} \left(e^{i {\bm q}\cdot {\bm x}}-1\right) 
D^{-1}_{S,S}({\bm q}) \; , 
\label{eq:Gphi-exp}
\end{equation}
where $D^{-1}_{S,S}({\bm q})$ is the diagonal sector of the  covariant derivative Green function given by (see Appendix 
\ref{Ap:GFD} for its computation),
\begin{equation} 
D^{-1}_{S,S}({\bm q})= \sum_{\ell} \frac{e^{i 2\pi\ell \left({\bm v}_S\cdot {\bm q}/\omega_c\right)}}{\hat {\bm n}_S\cdot {\bm q}} \; .
\label{eq:D-1SS}
\end{equation}
Replacing Eq. \eqref{eq:D-1SS} into Eq. \eqref{eq:Gphi-exp}, and integrating over $q_n$ and $q_t$ (in the local frame) we find, 
\begin{equation}
G_{\phi}(S,{\bm x})\!=
\begin{cases}
\!\sum_\ell \ln\Big[\frac{i/\lambda+2\pi\ell\ell_B^2k_F}{\hat {\bm n} \cdot {\bm x}+i/\lambda+2\pi\ell\ell_B^2k_F}\Big], & \textrm{for}\;   |\hat {\bm n}\times {\bm x}|\Lambda \ll 1  \\
\!-\infty,  &  \textrm{for} \; |\hat {\bm n} \times {\bm x}|\Lambda \gg1  
\end{cases}
\end{equation}
Plugging  this result into the equal-time propagator of the patch fermion, Eq.\eqref{eq:equaltime},  we find,
\begin{align}
\langle \psi^\dagger_S({\bm x})\psi_S(0)\rangle&= \frac{i\Lambda}{(2\pi)^2}
\prod_\ell \frac{1+i 2\pi\ell \ell_B^2k_F\lambda}{\hat {\bm n} \cdot {\bm x}+i/\lambda+2\pi\ell\ell_B^2k_F} 
\nonumber \\
&\sim \frac{i\Lambda}{(2\pi)^2} \sum_\ell  \frac{1}{\hat {\bm n} \cdot {\bm x}+2\pi\ell\ell_B^2k_F+i/\lambda} + 
\nonumber \\ 
&+O(1/N_L)
\end{align}
for $|\hat {\bm n}_S\times {\bm x}|\Lambda \ll1$ and zero otherwise. The last expression coincides with the result for the fermion propagator, 
Eq. \eqref{eq:Propagator-ellvarphi}, at equal times $r_0=0$, with corrections at least of order $N_L^{-1}$.

\section{Quantum dynamics of the Fermi surface in a magnetic field}
\label{Sec:BosDyn}

To describe the Fermi surface dynamics,  we write the generating functional  in a coherent state path integral representation,\cite{CastroNeto-1995}  
\begin{equation}
Z=\int \left(\prod_S{\cal D}a^\dagger_S{\cal D}a_S\right)\;  e^{i S[a_S,a^\dagger_S]} \; , 
\label{eq:PathItegral}
\end{equation}
where the action is
\begin{equation}
S[a_S,a^\dagger_S]=\sum_S\int dt d^2x \left\{a^\dagger_S i\partial_t a_S -H_B[a_S,a^\dagger_S]\right\}\; .
\end{equation}
Using the explicit form of $H_B$, Eq. \eqref{eq:BHa},  the action takes de form 
\begin{equation}
S[a]=\int d^2q d\omega \sum_{S,T} \; a^\dagger_S(\bm{q},\omega) M^{-1}_{S,T}({\bm q},\omega)a_T({\bm q},\omega) \; , 
\end{equation}
with
\begin{equation}
M^{-1}_{S,T}({\bm q},\omega)
= \left\{\omega-v_F \tilde D_S({\bm q}) \right\}\delta_{S,T}
\label{eq:M-1}
\end{equation}
and the covariant derivative,  $\tilde D_S({\bm q})$, was given in Eq. \eqref{eq:CovariantDerivative-q}.

The Bosonic correlation function, 
\begin{equation}
\langle a^{\dagger}_S(\omega,{\bm q}), a_T(\omega,{\bm q})\rangle= M(\omega,{\bm q}, \varphi_S,\varphi_T)\; , 
\end{equation}
is found  by solving the linear differential equation, 
\begin{equation}
\left(\omega-{\bm v}_S\cdot {\bm q}+i \omega_c\frac{\partial}{\partial \varphi_S} \right) M_{S,T}=\delta_p (\varphi_S-\varphi_T)\; , 
\end{equation}
where $\delta_p(\varphi_S)=\sum_n \delta(\varphi_S+2\pi n)$ is the periodic Dirac delta function.

We can express the Green function in terms of eigenvalues, $\lambda_n$, and eigenfunctions, $\psi_n(\omega,{\bm q},\varphi_S )$, of the operator $M^{-1}_{S,T}$. We have, 
\begin{equation}
M_{S,T}= \sum_n \frac{1}{\lambda_n} \psi^*_n(\varphi_S)\psi_n(\varphi_T)
\label{eq:MSTexpansion}
\end{equation}
provided $\lambda_n\neq 0$.
The eigenvalue equation reads, 
\begin{equation}
\left(\omega-{\bm v}_S\cdot {\bm q}+i \omega_c\frac{\partial}{\partial \varphi_S} \right)\psi_n=\lambda_n\psi_n\; , 
\end{equation}
with periodic boundary conditions $\psi_n(\varphi_S)=\psi_n(\varphi_S+2\pi)$. 
The solutions of this equation are the orthonormal eigenfunctions, 
\begin{equation}
\psi_n(\varphi_S)=\frac{1}{2\pi}e^{i n \varphi_S-i\frac{v_F q}{\omega_c}\sin\varphi_S} \; ,
\label{eq:eigenfunctions}
\end{equation}
and the eigenvalues are $\lambda_n=\omega- n\omega_c$, with $n$ integer.
By replacing Eq. \eqref{eq:eigenfunctions} into Eq. \eqref{eq:MSTexpansion}, we obtain
\begin{align}
M_{S,T}(\omega,{\bm q})&=\frac{1}{(2\pi)^2} e^{-i \frac{v_F q}{\omega_c}(\sin\varphi_S-\sin\varphi_T)} 
\nonumber \\
&\times 
\sum_{n=-\infty}^\infty \frac{e^{i n(\varphi_S-\varphi_T)}}{\omega-n\omega_c} \; .
\label{eq:MST}
\end{align}
with the desired property, $M_{S,T}=M^*_{T,S}$.
We observe that $M_{S,T}$ has poles at frequencies corresponding to the energies of the Landau levels, $\omega=n\omega_c$. 
To the best of our knowledge,  this is the first time that the Landau level quantization  appears in a bosonization procedure. 
An equivalent phenomenon has been described in the context of a hydrodynamic description of composite fermions in  a half-filled Landau level.\cite{Kim-1995}
Here,  the Landau levels appear as zero modes, $\lambda_n=0$, of the operator $M^{-1}_S=\omega- \tilde D_S({\bm q})$, upon imposing periodic boundary conditions around the Fermi surface. Thus, in our context, Landau quantization is a global property of  the Fermi surface,  and cannot be seen in a single patch expansion.

The limit of small frequencies, $\omega \ll v_F q$, and in particular  the limit $\omega\to 0$, is highly singular an needs to be handled with  care. To this end, it  is useful to rewrite the Green function in terms of the dual Fourier series of the Landau levels. Using the Poisson summation  formula, we have
\begin{equation}
\sum_{n=-\infty}^{\infty} \frac{e^{i n \theta}}{\omega-n\omega_c}=\sum_{\ell=-\infty}^{\infty}\int_{-\infty}^{\infty}
dz \frac{e^{i z (\theta-2\pi \ell)}}{\omega-z\omega_c}  \; .
\end{equation}
Defining $\theta=\varphi_S-\varphi_T$ as the angle subtended by two patches (see Fig. \ref{fig:patches}) and performing the $z$ integration (analytically continued to  the complex plane),  we find 
\begin{equation}
M(\omega,{\bm q},\varphi_S,\theta)=\frac{i}{2\omega_c} 
\sum_\ell \sgn(\theta-2\pi\ell)\exp(i\alpha_\ell) \; ,
\label{eq:MSTell}
\end{equation}
where we defined  the phase 
\begin{align}
\alpha_\ell(\omega,{\bm q},\varphi_S,\theta)&=\frac{\omega}{\omega_c}\left(\theta-2\pi \ell\right)
\label{eq:Mtheta} \\ 
&-
\frac {{\bm v}_S\cdot {\bm q}}{\omega_c} \sin\theta-
\frac {{\bm v}_S\times {\bm q}}{\omega_c} \left(1-\cos\theta\right) \; .
\nonumber
\end{align}
Eq. \eqref{eq:MSTell} is completely equivalent to Eq. \eqref{eq:MST}.

In order to compute the intra-patch (diagonal sector) Green function, we coarse-grain the bosonic Green function on each patch, by integrating over the  variable $\theta$ on one patch. We define
\begin{equation}
G_{SS}(\omega,{\bm q})= \int_{-\Lambda/2k_F}^{+\Lambda/2k_F} d\theta\; M(\omega,{\bm q},\varphi_S,\theta)  
\; .
\label{eq:GSS-theta}
\end{equation}
In the limit, $|\theta| \ll1$,    $\sin\theta\sim \theta$ and $1-\cos\theta\sim \theta^2$.  Thus, at leading order, the phase $\alpha_\ell$ becomes, 
\begin{equation}
\alpha_\ell(\omega,{\bm q},\varphi_S,\theta)=\frac{\omega}{\omega_c}\theta
-
\frac {{\bm v}_S\cdot {\bm q}}{\omega_c}\left(\theta+2\pi \ell\right)
\nonumber
\end{equation}
where we have shifted $\theta\to \theta+2\pi\ell$ in order to have a well-defined limit for small frequencies $\omega \ll {\bm v}_S\cdot {\bm q}$.  The result of the integration in Eq. \eqref{eq:GSS-theta} is, in the limit $N_L/N \gg 1$, 
\begin{equation}
G_{SS}(\omega,{\bm q})= \sum_\ell \frac{e^{-i 2\pi\ell\left(\frac{{\bm v}_S\cdot {\bm q}}{\omega_c}\right)}}{\omega-{\bm v}_S\cdot {\bm q}+i \Delta\epsilon \sgn(\omega)} \; , 
\label{eq:GSS}
\end{equation}
where we  have performed the analytic continuation $\omega\to \omega+ i\Delta \epsilon  \sgn(\omega)$.

The  $\ell=0$ term in Eq.\eqref{eq:GSS} is the well-known result for the particle-hole correlation function of a Fermi liquid. The effect of the magnetic field is to introduce oscillatory terms, which lead to the  discretization of the spectrum. This is depicted schematically by the shaded area in Fig. \ref{fig:patches}.
Finally, we note that this expression has the correct low frequency behavior since
\begin{equation}
\lim_{\omega\to0} G_{SS}(\omega,{\bm q})=-D^{-1}_{SS}({\bm q})\; ,
\end{equation}
used above in Eq. \eqref{eq:D-1SS}. An independent detailed computation is presented in Appendix \ref{Ap:GFD}. 

In the same way we can compute the inter-patch (off-diagonal) bosonic correlation functions. Consider $\theta=\bar \theta+\delta\theta$, where $\bar\theta$ is the finite angle subtended by the patches $S$ and $T$, and integrate over $|\delta\theta|<\Lambda/k_F$.  The result is 
\begin{align}
G_{ST}&(\omega,{\bm q})=\nonumber\\
 \sum_\ell 
& \frac{\sin\left[ \frac{\Lambda}{2k_F\omega_c} (\omega-{\bm v}_S\cdot {\bm q} \cos\bar\theta+\bm{ v}_S\times\bm{q} \sin\bar\theta)    \right] }
{\omega-\bm{v}_S\cdot\bm{q} \cos\bar\theta+\bm{v}_S\times\bm{q} \sin\bar\theta+i \Delta\epsilon \sgn(\omega)} e^{-i\alpha_\ell(\bar\theta)} \; ,
\end{align}
with $|\bar\theta|>\Lambda/k_F$. For weak magnetic fields, this a strongly oscillating function of $\omega$. Due to the analytic continuation $\omega\to \omega+i\Delta\epsilon \sgn(\omega)$, the Fourier transform  yields an exponential factor of the form 
$G_{S,T}(t,{\bm x})\sim \exp(-N_L/N)$, that enforces the condition that the limit $\lim_{\omega_c\to 0}G_{S,T}(t,{\bm x})=0$, as it should be.

\section{de Haas-van Alphen oscillations}
\label{Sec:dHVA}

A distinct characteristic of metals in weak magnetic fields is the presence of quantum oscillations.\cite{Schubnikov-1930,deHaas-1930}
They encode the main features and existence of a Fermi surface.
The de Haas-van Alphen effect (dHvA),\cite{deHaas-1930,kosevich-1956-2} consists of the periodic oscillation of the magnetization as a function of an external magnetic field. It is widely used as a tool to investigate  the structure and topology of the Fermi surface. In this section we show how the dHvA  can be computed by using the bosonization approach.
In particular, we show that the structure of the oscillations is contained in the bosonized action, and  depends on the dynamics of the collective modes of the Fermi surface.  This is important because, on the one hand, confirms that the bosonization procedure we present here  correctly captures the structural properties of the Fermi surface. On the other hand, it opens the possibility of computing quantum oscillations in extensions of the present work to strongly correlated systems, e.g. the nematic Fermi fluid. 

We begin with the bosonic path integral representation of  Eq. \eqref{eq:PathItegral}. Integrating over the bosonic fields $a_S$ and $a^\dagger_S$ we find the zero temperature limit of the thermodynamic potential, $F=-\ln Z$. We find
\begin{equation}
F(\omega_c)=-{\rm Tr}\ln \left(M^{-1}\right)  \; , 
\label{eq:Fomega}
\end{equation}
where the operator $M^{-1}$ is the kernel given by Eq. \eqref{eq:M-1}.

The orbital magnetic moment, $\mathcal{M}$, can be computed as
\begin{align}
{\cal M}&=-\left(\frac{\mu_B}{L^2}\right)\frac{\partial F}{\partial B}=-\mu_B\left(\frac{v_F}{k_F L^2}\right)\frac{\partial F}{\partial \omega_c} \nonumber \\
&=\mu_B\left(\frac{v_F}{k_F L^2}\right){\rm Tr} \left\{M \; \frac{\partial M^{-1}}{\partial \omega_c}\right\} \; .
\end{align}
Here $\mu_B=1/2m=v_F/2k_F$ is the Bohr magneton in the natural units for our system.
The operators $M$ and $M^{-1}$ are diagonal in both  frequency and momentum variables $(\omega,{\bm q})$. However, they are not diagonal in 
the patch index $S$. Then, the trace can be computed as
\begin{align}
{\rm Tr} \Big\{M & \times\frac{\partial M^{-1}}{\partial \omega_c}\Big\}
= \nonumber \\
&\frac{L^2}{v_F\lambda}\int d\omega d^2 q
\sum_{S,T}\langle S| M |T \rangle\langle T|\frac{\partial M^{-1}}{\partial \omega_c}|S\rangle  \; .
\end{align}
Here, for simplicity, we use the notation $|S\rangle \equiv |S,\omega,{\bm q}\rangle$.

On the other hand, 
\begin{equation}
\langle T|\frac{\partial M^{-1}}{\partial \omega_c}|S\rangle=i \frac{\partial}{\partial \varphi_S} \delta(\varphi_S-\varphi_T).
\end{equation}
 Then the magnetization takes the form, 
\begin{equation}
{\cal M}=i \left(\frac{\mu_B}{k_F\lambda}\right)
\int d\omega d^2 q
\sum_{S} \left. \frac{\partial M_{S,T}}{\partial\varphi_T}\right|_{T=S} \; .
\end{equation}
Here, the condition $T=S$ should be understood as $\varphi_T=\varphi_S$ in the sense of a coarse-grained expression on a patch, 
{\em i.\ e.\ },   $-\Lambda/2k_F<\varphi_S-\varphi_T<\Lambda/2k_F$. Thus, using $\theta=\varphi_S-\varphi_T$ we have, 
\begin{equation}
\left.\frac{\partial M_{S,T}}{\partial\varphi_T}\right|_{T=S} =-\lim_{\Lambda/k_F\to 0}\int_{-\Lambda/2k_F}^{\Lambda/2k_F} d\theta \frac{d M(\varphi_S, \theta)}
{d\theta} \; ,
\end{equation}
where $M(\varphi_S, \theta)$ is given by Eq. \eqref{eq:MSTell}.

Now, we choose a local coordinate system in each patch, $d^2q=dq_n dq_t$, and integrate over $q_t$ and $\omega$. After taking the continuum  limit $(\Lambda/k_F)\sum_S\to \int_0^{2\pi}d\varphi_S$, we find the following expression for the orbital magnetization $\mathcal{M}$
\begin{align}
{\cal M}&=-\frac{1}{2}\left(\frac{v_F^2}{k_F\omega_c}\right)\sum_\ell\int_0^\lambda
 dq \int_0^{2\pi}d\varphi_S\;  e^{i2\pi\ell \frac{v_F q}{\omega_c} \cos\varphi_S} \nonumber \\
 &=-\left(\frac{\pi v_F^2}{k_F\omega_c}\right) \sum_\ell\int_0^\lambda
 dq   J_0(2\pi\ell v_F q/\omega_c) \; ,
\end{align}
where $J_0(z)$ is the Bessel Function of the first kind. 
Finally, we compute the $q$ integration, remembering that we are working in  the asymptotic  limit $v_F\lambda/\omega_c\gg 1$. Disregarding  constant terms, we find for the oscillatory part of the magnetization  the result
\begin{equation}
{\cal M}_{\rm osc}=-2\mu_B \sqrt{\frac{\omega_c}{2\pi v_F\lambda}} \sum_{\ell=1} \frac{1}{\ell^{3/2}}\sin\left\{\frac{ 2\pi\ell v_F \lambda }{\omega_c}+\frac{\pi}{4}\right\} \; .
\label{eq:Osc}
\end{equation}
We see that, the magnetization oscillates with the variable $1/\omega_c$, with period $1/(\ell v_F\lambda)$. Each harmonic is weakly damped by the power law $\ell^{-3/2}$. The overall oscillating function is damped at low fields with an envelope  function $(\omega_c/2\pi v_F\lambda)^{1/2}$. This result coincides with the zero-temperature limit of the well-known Lifshitz-Kosevich formula.\cite{kosevich-1956-2}
From a geometrical point of view, the envelope function is the flux of the magnetic field through the area $1/2\pi k_F\lambda$,  corresponding (in momentum space) to the  area  limited  by the shell of width $\lambda$ around the Fermi surface. It is  interesting to note that the argument of the $\sin$ function, $v_F\lambda/\omega_c$, is the number of Landau levels contained in each patch. Thus, the magnetization completes one period every time an additional Landau level enters the patch, as the magnetic field is lowered. 

\begin{figure}[ht]
\includegraphics[scale=0.8]{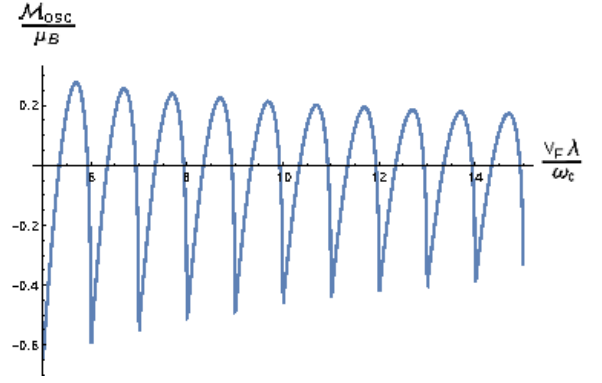}
\caption{Orbital magnetization ${\cal M}_{\rm osc}/\mu_B$, Eq. (\ref{eq:Osc}), as a function of $v_F\lambda/\omega_c$. 
We have considered the first twenty terms in the sum. Higher harmonics make no difference within the graphic precision. Each time the variable 
 $v_F\lambda/\omega_c$ is an integer, the magnetization has a cusp, corresponding with an additional complete filled Landau level inside the patch. }
\label{fig:Oscillations}
\end{figure}

In figure \ref{fig:Oscillations} we depict the orbital magnetization, Eq. (\ref{eq:Osc}), where we have summed the first twenty harmonics.  Higher harmonics make no difference within the graphic precision.
\begin{figure}[ht]
\includegraphics[scale=0.8]{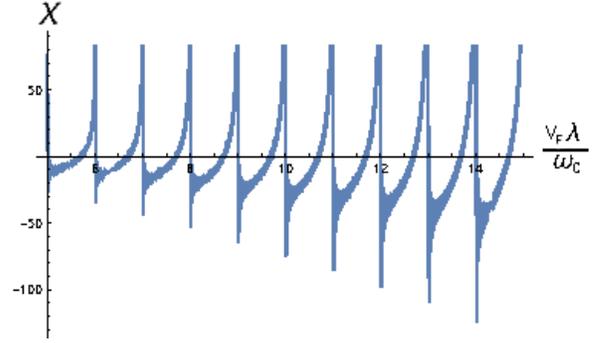}
\caption{Magnetic susceptibility $\partial{\cal M}_{\rm osc}/\partial B$, as a function of $v_F\lambda/\omega_c$. 
We have considered the first twenty terms in the sum.  
For $v_F\lambda/\omega_c=$integer, the susceptibility diverges, corresponding with the inclusion on an additional complete filled Landau level inside the patch.
}
\label{fig:Susceptibility}
\end{figure}
In fig. \ref{fig:Susceptibility} we  present
the magnetic susceptibility $\chi=\partial {\cal M}/\partial B$. We observe divergencies, each time a new Landau level enters the region bounded by 
$|{\bm q}|\leq\lambda/2$. The divergence arises because higher harmonics are damped with the extremely low power $\ell^{-1/2}$, producing a divergent series at each Landau level. Even though, this is a zero temperature computation,  temperature fluctuations put another energy scale in the problem. For relatively strong fields, $\omega_c>k_B T$ (where $k_B$ is the Boltzmann constant), the results essentially coincide with the zero temperature expression. For weak fields, $\omega_c<k_B T$,  temperature exponentially damps out higher harmonics, and only the leading term, $\ell=1$, survives. In this case,  the divergencies of the susceptibility are rounded by temperature. 

\section{Forward scattering interactions}
\label{Sec:Interactions}
The power of bosonization as a  non-perturbative technique is due in part to the fact that two-body forward scattering interactions 
can be written as quadratic operators in the bosonic basis.
The more general two-body  interaction term in the low-energy Hamiltonian can be written in terms of products of  four fermionic operators, $\psi^\dagger_{S_1}({\bm x})\psi^\dagger_{S_2}({\bm y})\psi_{S_3}({\bm x})\psi_{S_4}({\bm y})$, 
where $\{S_1,S_2,S_3,S_4\}$ is a set of patches over the Fermi surface. 
Symmetries strongly constrain the possible terms that should  be considered.
Momentum conservation, ${\bm k}_{S_1}+{\bm k}_{S_2}+{\bm k}_{S_3}+{\bm k}_{S_4}=0$, selects two kinds of processes: forward scattering interactions for which, $S_1=S_3$ and $S_2=S_4$, and BCS type interactions for which $S_1=-S_2$ and $S_3=-S_4$. Moreover, exchange processes ($S_1=S_4$, $S_2=S_3$ ) are absent in the spinless model with reasonably local interactions. 
In the absence of magnetic fields, these types of low-energy Fermi surface interactions were extensively studied both in the fermionic basis,\cite{Shankar1994,Metzner-1998} as well as with bosonization techniques.\cite{CastroNeto-1993,CastroNeto-1994,CastroNeto-1995,houghton-1993,houghton-1994,haldane-1994,houghton-2000,BaOx2003,Lawler2006,lawler-2007}
 
Considering a simple model,  based on a circular Fermi surface,  couplings between different patches can be classified according with irreducible representations of the rotation group. Each representation is labeled with an angular momentum quantum number $\ell$. Then, it is possible to consider forward-scattering interactions between charge densities ($\ell=0$), dipole  moment density ($\ell=1$),  quadrupole moment density ($\ell=2$), and so on. The simplest term is a density-density interaction of the form, 
\begin{equation}
H_{\rm int}= \sum_{S,T}\int d^2xd^2y f_0({\bm x}-{\bm y}) \psi^\dagger_S({\bm x})\psi_S({\bm x})\psi^\dagger_T({\bm y})\psi_T({\bm y})
\label{eq:HintF0}
\end{equation} 
where, in the Fermi liquid phase, the interaction potential can be  considered as being essentially local, $f_0({\bm x}-{\bm y})=f_0\delta^2({\bm x}-{\bm y})$.

By carefully computing the point-splitting product of fermion operators, $\lim_{\epsilon\to 0}:\psi_S^{\dagger}({\bm x}+\epsilon {\bm n})\psi_S({\bm x}-\epsilon {\bm n}):$, using the fermion operator defined in  Eq. (\ref{eq:Fermion-Operator}), we find the usual bosonization rule 
\begin{equation}
\psi_S^\dagger({\bm x})\psi_S({\bm x})\to (1/L)\;\delta n_S({\bm x}).
\end{equation}
Notice, however, that the bosonization of the {\em local charge density operator} $\rho({\bm x})=\psi^\dagger({\bm x}) \psi({\bm x})$ contains, in addition to this contribution, other operators  involving fermion operators on different patches, e.g. the charge-density-wave operator that involves patches on opposite sides of the Fermi surface. We will not consider these interesting effects in this work.

With this identification, the interaction Hamiltonian, Eq. (\ref{eq:HintF0}) is bosonized to a quadratic operator
\begin{equation}
H_{\rm int}=\frac{f_0}{L^2}\int d^2x \sum_{S,T} :\delta n_S({\bm x})\delta n_T({\bm x}): \; .
\label{eq:HintF0Bos}
\end{equation} 
Thus, the interacting fermionic model is mapped into a  system of non-trivially coupled harmonic oscillators. 

The bosonization of forward scattering interactions with higher angular momentum, such as, for instance, quadrupolar interactions, is more involved. The reason is that the quadrupole density contains second order covariant derivatives that do not commute. Thus, in order to bosonize them,  it is necessary to carefully implement a point-splitting computation on the Fermi surface variables. We postpone this study for a separate future presentation.  
We finish this section by showing a calculation of collective modes in the simpler, but non-trivial, case of density-density interactions.  

\subsection{Collective modes}

An instructive application of the bosonization technique in a magnetic field is the  computation of  Fermi surface collective modes. Let us consider the interacting Hamiltonian 
\begin{equation}
H=H_0+H_{\rm int}
\end{equation}
where $H_0$ and $H_{\rm int}$ are given by Eqs. (\ref{eq:Bosonic-Hamiltonian}) and (\ref{eq:HintF0Bos}) respectively. 
Using the covariant Schwinger algebra, Eq. (\ref{eq:Schwinger-algebra}), we can write the Heisenberg equation of motion, $\partial_t\delta n_S=-i[H,\delta n_S]$, as
\begin{align}
&\frac{\partial \delta n_S(\bm{x})}{\partial t}+{\bm v}_S\cdot {\bm \nabla}\delta n_S(\bm{x})-\omega_c \frac{\partial \delta n_S(\bm{x})}{\partial \varphi_S}\nonumber \\
&+F_0{\bm v}_S\cdot {\bm \nabla}\sum_T\delta n_T(\bm{x})
=0 \; , 
\end{align} 
where we  defined the dimensionless parameter $F_0=N(0) f_0$, where $N(0)=k_F/v_F$ is the one-particle density of states at the Fermi surface. 
Fourier transforming in ${\bm x}$ and $t$ we find, 
\begin{align}
\left(\omega-{\bm v}_S\cdot {\bm q}+i \omega_c\frac{\partial}{\partial \varphi_S} \right) \delta n_S=F_0\left({\bm v}_S\cdot {\bm q}\right)\rho({\bm q},\omega)
\end{align} 
where we have introduced the (long wavelength) density $\rho(\bm{q},\omega)= \sum_S \delta n_S({\bm q},\omega)$.
This equation can be inverted as, 
\begin{equation}
\delta n_S({\bm q,\omega})=\rho({\bm q}, \omega)F_0\sum_T M_{S,T}\; {\bm v}_T\cdot {\bm q} \; ,
\label{eq:nF0rho}
\end{equation}
where the Green function $M_{S,T}$ is defined in Eq. (\ref{eq:MST}),  or equivalently,  Eq. (\ref{eq:MSTell}). This is a formal solution, since 
$\delta n_S$ is present on both sides of the equation, thorough $\rho({\bm q}, \omega)$. In order to find the self-consistent condition for the collective modes, we sum up Eq. \eqref{eq:nF0rho} over the patch variable $S$, and we take the continuum limit. Thus,  we find 
\begin{equation}
\frac{v_Fq}{(2\pi)^2}\int_0^{2\pi} d\varphi_S d\varphi_T\; M(\varphi_S,\varphi_T) \cos\varphi_T=\frac{1}{F_0}\; .
\end{equation}
For weak magnetic fields, the Green function $M(\varphi_S,\varphi_T)$ is strongly picked at $\varphi_S\sim\varphi_T$. Then, in that limit,  the integral over $\varphi_T$ can be easily done. At leading order in the magnetic field we  obtain, 
\begin{equation}
\sum_{\ell=-\infty}^{\infty}\int_0^{2\pi} \frac{d\varphi_S}{2\pi} \frac{\cos\varphi_S}{s-\cos\varphi_S}\;e^{i 2\pi\ell \frac{v_F q}{\omega_c}\cos\varphi_S}=\frac{1}{F_0}
\label{eq:collectivemodes}
\end{equation}
where we have introduced the usual Fermi liquid parameter $s=\omega/(v_F q)$. Solutions of Eq. \eqref{eq:collectivemodes} provide the dispersion relation $\omega({\bm q})$ of the collective modes of the system. The term with  $\ell=0$ is the zero magnetic field condition, found in usual Fermi liquids. The presence of a weak magnetic field produces highly oscillatory components for $\ell\neq 0$. In order to compute corrections to the Fermi liquid result we need to  integrate over $\varphi_S$. This calculation  can be done exactly using the  expansion\cite{gradshteyn2007}
\begin{equation}
e^{i z\cos\varphi}=\sum_{k=-\infty}^{+\infty} i^k J_k(z) e^{ik\varphi} \; 
\label{eq:expansion}
\end{equation}
where $J_k(z)$ is the $k^{\rm th}$ order Bessel function of the first kind. 
By replacing Eq. \eqref{eq:expansion} into Eq. \eqref{eq:collectivemodes}, calculating the angular integrals over $\varphi_S$, and taking advantage of the symmetry properties of the bessel functions, $J_{2k}(z)=J_{-2k}(z)$, $J_{2k+1}(z)=-J_{-(2k+1)}(z)$, we find that Eq.\eqref{eq:collectivemodes} takes the form
\begin{align}
&\chi_0(s)\left\{1+2 \sum_{\ell=1}^{\infty}J_0\left(2\pi\ell\frac{v_Fq}{\omega_c}\right)\right\}\nonumber\\
&+4\sum_{n=1}^{\infty} \chi_{2n}(s)\sum_{\ell=1}^{\infty} J_{2n}\left(2\pi\ell\frac{v_Fq}{\omega_c}\right) =\frac{1}{F_0}
\label{eq:collectivemodes1}
\end{align}
where. $\chi_n(s)$ are multipole Fermi liquid susceptibilities 
\begin{align}
\chi_n(s)&=\int_0^{2\pi}\frac{d\varphi}{(2\pi)}\frac{\cos\varphi}{s-\cos\varphi}e^{in\varphi} 
\end{align}
When needed, we have analytically continued  the variable $s\to s+i\epsilon \sgn(s)$. 
For $s>1$, \begin{equation}
\chi_n(s)=-\delta_{k,0}+\frac{s}{\sqrt{s^2-1}} \left(s-\sqrt{s^2-1} \right)^n.
\end{equation}

In the second line of  Eq. (\ref{eq:collectivemodes1}), the sum over $n$ can be exactly performed in  the asymptotic limit $v_F q/\omega_c\gg 1$. 
We have found the following expression, 
\begin{align}
&\sum_{n=1}^{\infty} \chi_{2n}(s) J_{2n}\left(2\pi\ell\frac{v_Fq}{\omega_c}\right)= \nonumber \\
&-(1+\chi_0(s))\frac{(s-\sqrt{s^2-1})^2}{1+(s-\sqrt{s^2-1})^2} J_0\left(2\pi\ell\frac{v_Fq}{\omega_c}\right)
\nonumber \\
&+O\left[\left(\frac{\omega_c}{v_F q}\right)^{3/2}\right]
\end{align}
Replacing this result into Eq. \eqref{eq:collectivemodes1}, we finally get 
\begin{align}
\chi_0(s)\left\{1+4\frac{\sqrt{s^2-1}}{s}\sum_{\ell=1}^{\infty}J_0\left(2\pi\ell\frac{v_Fq}{\omega_c}\right)\right\}
=\frac{1}{F_0}
\label{eq:collectivemodes2}
\end{align}
This is a non-trivial algebraic equation for $\omega({\bm q})$. In fact it is a non-analytic equation in terms of the two dimensionless variables, 
$s=\omega/v_F q$ and $\omega_c/v_F q$.  
We have solved Eq.\eqref{eq:collectivemodes2} perturbatively in the parameter $(\omega_c/v_F q)^{1/2}$, considering $s>1$, and obtained, 
\begin{align}
s&=\frac{1+F_0}{\sqrt{(1+F_0)^2-F_0^2}}
\nonumber \\
&+
\Delta(F_0)\left(\frac{\omega_c}{v_F q}\right)^{1/2}
\sum_{\ell=1}^{\infty} \frac{e^{-2\pi\ell \epsilon/\omega_c}}{\sqrt{\ell}} \cos\left(\frac{2\pi\ell v_F q}{\omega_c}-\frac{\pi}{4}\right)
\nonumber \\
&+O\left[\left(\frac{\omega_c}{v_F q}\right)^{3/2}\right]\; , 
\label{eq:zerosound}
\end{align}
where we have define the quantity
\begin{equation}
\Delta(F_0)=\frac{4}{\pi}\frac{F_0^3}{(F_0+1)[(1+F_0)^2-F_0^2]^{3/2}}  \; .
\label{eq:DeltaF0}
\end{equation}

\begin{figure}
  \begin{subfigure}{8 cm}
\includegraphics[width=0.45\textwidth]{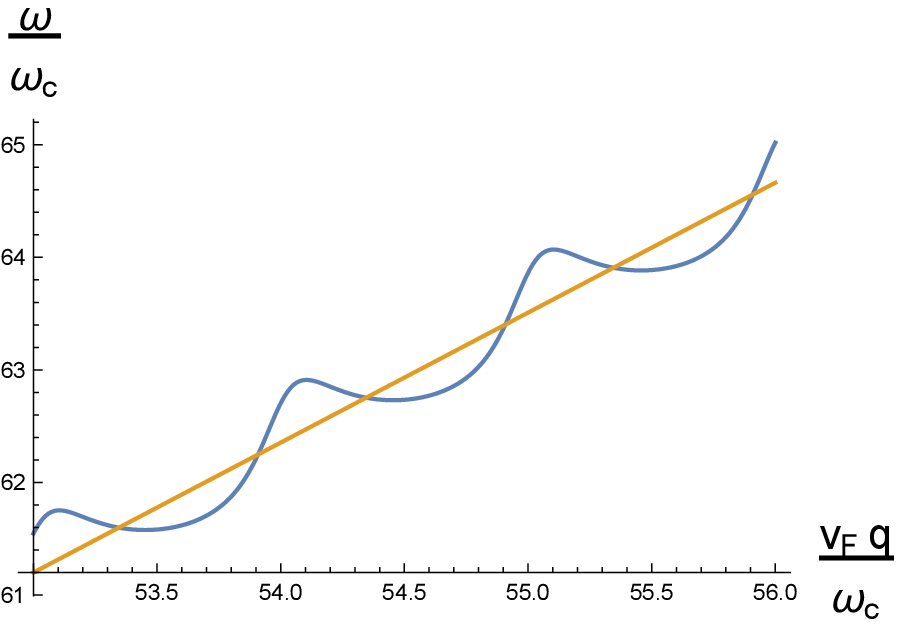}
  \end{subfigure}
  \begin{subfigure}{8 cm}
    \centering
\includegraphics[width=0.45\textwidth]{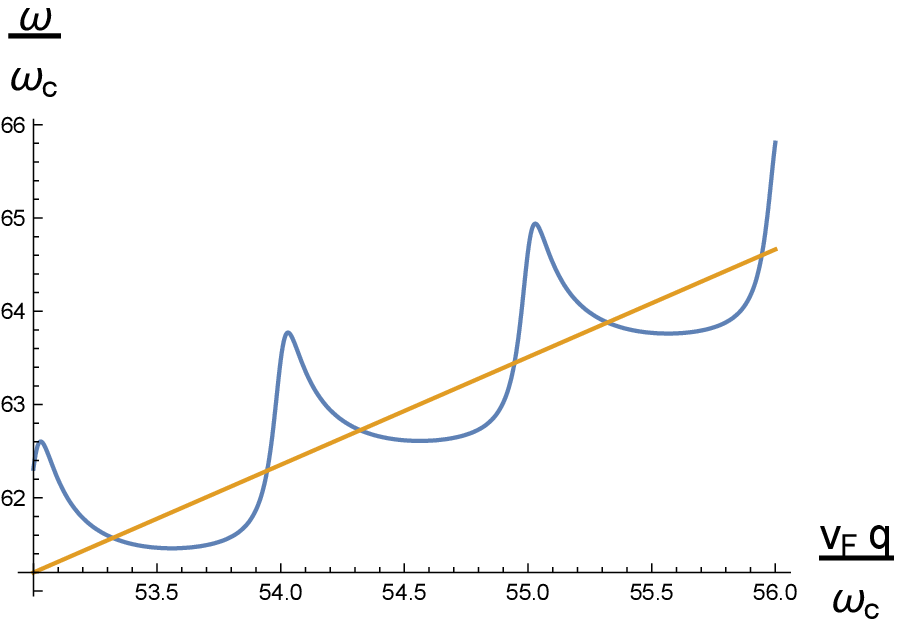}
  \end{subfigure}
 \caption{Dispersion relation for the collective mode described by Eq. (\ref{eq:zerosound}). In both figures we have fixed $F_0=1$. The straight line is the classical Landau result for the zero sound mode. In (a), we have fixed  $\epsilon/\omega_c=0.15$, corresponding to a weak magnetic field regime ($\omega_c< k_B T_c$). In (b), we fixed  $\epsilon/\omega_c=0.05$, corresponding with stronger values of the magnetic field ($\omega_c\gtrsim k_B T_c$). The peaks coincide with values of 
$v_F q/\omega_c$ integers.}
\label{fig:zerosound}
\end{figure}

Equation (\ref{eq:zerosound}) is the main result of this section. It represent the corrections to the zero sound dispersion relation at leading order in $(\omega_c/v_Fq)^{1/2}$.
Taking the limit $\omega_c\to 0$ in Eq. (\ref{eq:zerosound}), only survives the first line of the equation, {\em i.\ e.\ }, 
\begin{equation}
s=\frac{1+F_0}{\sqrt{(1+F_0)^2-F_0^2}} \; ,
\label{eq:s0}
\end{equation}
which is the well known   linear dispersion relation of the underdamped Landau zero sound in two dimensions.\cite{CastroNeto-1994}
The second line in Eq. (\ref{eq:zerosound}), is the leading order correction due to the presence of the homogeneous magnetic field. The corrections are highly oscillatory terms whose periods are integer multiples of  $\omega_c/\ell$.    
The damping factor in the sum over  $\ell$ appears due to the analytic continuation $\omega\to \omega+i\epsilon \sgn(\omega)$, necessary to properly define the Green function. In the presence of a magnetic field, it is subtle to take the limits $\omega_c\to 0$ and $\epsilon\to 0$. If we take $\epsilon\to 0$ at fixed magnetic field, all harmonics contribute to the sum, producing singularities each time the energy of the particle-hole pair, $v_F q$, which coincides with the energy of a Landau level. Conversely, if we take the limit $\omega_c\to 0$ first, all harmonics are damped to zero. In practice, the scale of $\epsilon$ is given by temperature, that blurs the finely discretized dense spectrum. Thus, if $\omega_c\lesssim k_B T$, only very few harmonics  contribute to the sum, producing a nicely oscillatory behavior. On the other hand, if $\omega_c\gtrsim k_B T$, more and more harmonics contribute to the sum, producing peaks whenever a Landau level coincides with the energy of the particle-hole pair. In the extreme case of  $\omega_c\gg k_B T$, equivalent to zero temperature,  the result turns out to be discontinuous peaks. However, in this regime, the gaps are important and the Fermi liquid type description breaks down. 
We depict this behavior in figure \ref{fig:zerosound}. Using Eq. (\ref{eq:zerosound}), where we have plotted 
$\omega/\omega_c$, as a function of the dimensionless variable $v_F q/\omega_c$. The straight line is the classical result of Landau, for a typical value of $F_0=1$. In  \ref{fig:zerosound}a, we fixed the damping term $\epsilon/\omega_c= 0.15 $. In this case, only the very few first harmonics appreciably contribute to the sum, producing the oscillatory behavior shown in the figure.    
In Fig. \ref{fig:zerosound}b, we fixed the damping term one order of magnitude weaker, $\epsilon/\omega_c=0.05$,  corresponding to a regime with stronger magnetic fields. We observe pronounced peaks  for integer values of $v_F q/\omega_c$, {\em i.\ e.\ }, when the energy of the particle hole excitation is commensurate with the energy of a Landau level.

\begin{figure}
\includegraphics[scale=0.8]{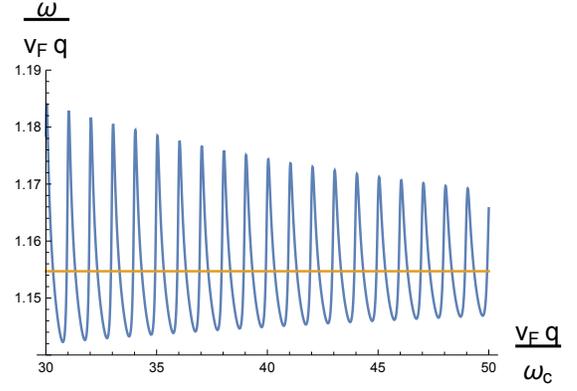}
\caption{Energy of the collective mode excitation in units of $v_F q$, for fixed $q$, as a function of the inverse magnetic field in the same scale, $v_F q/\omega_c$. The curve was depicted with Eq. (\ref{eq:zerosound}) where we have fixed $F_0=1$ and  $\epsilon/v_F q=0.002$. The horizontal line is the value of the zero sound energy in the absence of magnetic field,  given by Eq. (\ref{eq:s0}).}
\label{fig:W-oscillations}
\end{figure}

In Fig. \ref{fig:W-oscillations} we have fixed the value of the particle-hole excitation energy $v_Fq$, and have depicted the frequency of the collective mode in terms of the inverse magnetic field $\omega_c^{-1}$, in the same scale, using Eq. (\ref{eq:zerosound}).  To draw the picture we  chose
$F_0=1$ and $\epsilon/v_F q=0.002$. We observe a damped oscillation as we decrease the magnetic field. 
This behavior is another manifestation of quantum oscillations. However, different form the dHvA effect, it is proper of a Fermi liquid and it is not present in the Fermi gas. Indeed, the amplitude of the oscillations is governed by $\Delta(F_0)$.  As we can see from Eq. (\ref{eq:DeltaF0}),  
$\lim_{F_0\to 0}\Delta(F_0)=0$ and, in this case, there is no correction to the free dispersion,  $\omega=v_F q$. 

 \section{Discussion and Conclusions}
\label{Sec:Conclusions}

In this paper we have presented a bosonization technique that describes a 2D system of fermions at fixed density in  a weak magnetic field. 
We focused on a regime in which there is a large number of Landau levels  in a small energy band around the chemical potential. In this way, the system is, in some sense, near a Fermi liquid regime, since the concept of Fermi surface still make sense.
The main idea of the bosonization approach is to project the states of the system onto a restricted subspace, spanned by states very near the Fermi surface. At very low temperature, it is assumed that fluctuations of the Fermi surface are responsible for  the main properties of a fermionic system. Processes originating  from transitions deep inside the Fermi sea are completely irrelevant in this low energy regime. 
Within this perspective,  we  coarse-grained the fermion operator in  small patches of the Fermi surface, in which we can linearize the dispersion relation. Fermi surface deformations can be parametrized by a set of bosonic operators describing the creation of particle-hole excitations  in each patch.

We  showed that these operators satisfy a magnetic-field-dependent {\em covariant Schwinger algebra}. In addition to the usual Schwinger term, proportional to the normal derivative of the delta function, there is now a tangential term, proportional to the magnetic field,   that mixes neighbor patches. 
This algebra is one of the important results of the paper, and it is the cornerstone of bosonization construction.
The bosonization program is completed by constructing the full fermion operator in terms of bosonic excitations. Unlike   the case of bosonization of the  Fermi liquid, the phase of a fermionic operator on each patch depends on a coherent superposition of bosonic fields all around the Fermi surface. With this construction at hand, we  computed the fermionic equal time propagator, showing that it coincides with the one computed with standard procedures. Moreover,  we have  shown that the low-energy fermionic Hamiltonian can be exactly mapped to  a local quadratic bosonic Hamiltonian.    

The dynamics of the Fermi surface can be represented by a path integral coherent state formalism. The effective action is written in terms of 
a set of couple canonical bosonic fields. We  showed how to use this formalism to compute bosonic correlation functions. As an example, we have computed the orbital magnetization and   showed that the bosonized action correctly captures quantum oscillations responsible for the de Haas-van Alphen effect. The fact that quantum oscillations can be obtained from the bosonized action opens the  possibility of computing them at strong coupling. 
We have also studied the bosonization of forward scattering interactions, encoded in the Landau parameter $F_0$. We  computed the spectrum of collective modes, in particular, corrections to the Landau zero sound mode.  We have shown that the corrections are damped oscillations, in terms of the momentum, and either in term of the inverse magnetic field. These results resemble the problem of quantum oscillations and, in fact, have the same origin. However, this oscillation, in a non-equilibrium property, is completely due to interactions. In fact, the correction vanishes in the limit of $F_0\to 0$. 

In this work we considered the case of a fluid of  spinless fermions at finite density  with the main goal to establish the bosonization rules for a system in a uniform magnetic field. We have also discussed some effects of  Fermi liquid corrections due to  forward scattering interactions.  An important and  challenging extension  is to apply these ideas and techniques to the more interesting case of 2D Fermi systems at and near quantum criticality. Higher dimensional bosonization was used before to study the quantum phase transition to a nematic state driven by a Pomeranchuk instability.\cite{lawler-2007} Such systems are the simplest examples of non-Fermi liquids. In a separate publication we will study the effects of a uniform magnetic field in these non-Fermi liquid phases of matter and quantum critical points. 
The great interest of such studies is how non-Fermi liquid behavior affects quantum oscillations and similar magnetic field driven phenomena.

\textit{Note}: After this work was completed we became aware of a recent paper by Nguyen and Son \cite{Nguyen-2018} on an algebraic approach to the fractional Hall effect. Although the problem that these authors study is formally different than ours, their formalism is in spirit similar to the theory of higher dimensional bosonization of dense Fermi  systems that we use  here.

\acknowledgments
The Brazilian agencies {\em Conselho Nacional de Desenvolvimento Cient\'\i fico e Tecnol\'ogico} (CNPq), {\em Funda\c c\~ao  Carlos Chagas Filho de Amparo \`a Pesquisa do Estado do Rio de Janeiro} (FAPERJ), and {\em Coordena\c c\~ao de Aperfei\c coamento de Pessoal de N\'\i vel Superior} (CAPES) are acknowledged for partial financial support. L.R. acknowledge the Institute of Condensed Matter Theory of UIUC for kind hospitality during his ``CAPES sandwich doctoral fellow".     D.G.B also acknowledges partial financial support by the  Associate Program of the Abdus Salam International Centre for Theoretical Physics, ICTP, Trieste, Italy. E.F. acknowledges
support from the US National Science Foundation through grant No. DMR 1725401 at the
University of Illinois.

\appendix
\section{The Fermion Propagator in a uniform magnetic field}
\label{Ap:Propagator}

The propagator for two-dimensional spinless free fermions in a uniform magnetic field, with $p$ filled Landau levels,  can be written as,\cite{lopez-1991}
\begin{align}
i G({\bm x},{\bm y}, t-t')=&
\nonumber \\
\Theta(t-t')\sum_{m=p}^\infty & \int\frac{dk}{2\pi} e^{-i\omega_m(t-t')}\varphi_{mk}({\bm x})\varphi^*_{mk}({\bm y})\nonumber \\
-\Theta(t'-t)\sum_{m=0}^{p-1} & \int\frac{dk}{2\pi} e^{-i\omega_m(t-t')}\varphi_{mk}({\bm x})\varphi^*_{mk}({\bm y})  , 
\end{align}
where $\omega_m=\omega_c(m+1/2)$ and the eigenfunctions of the Hamiltonian of Eq. \eqref{eq:Hamiltonian}, in the Landau gauge $A_1=-Bx_2, A_2=0$, are given by
\begin{align}
\varphi_{mk}({\bm x})=&\sqrt{\frac{1}{2^m m! \sqrt{\pi}\ell_B}} \nonumber \\
&\times e^{ik x_1} e^{-\frac{1}{2}\left(\frac{x_2}{\ell_B}-k\ell_B \right)^2}
H_m\left(\frac{x_2}{\ell_B}-k\ell_B\right)  ,
\end{align}
where $H_m$ are the Hermite polynomials of order $m$.

Integrating over $k$ we find, 
\begin{align}
i G&({\bm x},{\bm y}, t-t')=\frac{1}{2\pi\ell_B^2} e^{-\left(\frac{r}{2\ell_B}\right)^2} e^{i\theta_B}  \nonumber \\
&\times\left\{\Theta(t-t')\sum_{m=p}^\infty e^{-i\omega_m(t-t')} L_m[\left(r/2\ell_B\right)^2]\right.  \nonumber \\
&-\left. \Theta(t'-t)\sum_{m=0}^{p-1} e^{-i\omega_m(t-t')} L_m[\left(r/2\ell_B\right)^2]\right\}  ,
\end{align}
where  $r=|{\bm x}-{\bm y}|$, $r_0=t_x-t_y$,  
\begin{equation}
\theta_B({\bm x}, {\bm y})=
\frac{1}{2}B (x_1-y_1)(x_2+y_2) 
\end{equation}
with ${\bm x}=(x_1,x_2)$, ${\bm y}=(y_1,y_2)$, and $L_m$ are the Laguerre polynomials.

In the  weak magnetic field regime,  the number of filled Landau levels is huge, $p\gg1$, and  $\epsilon_F=k_F^2/2M\sim\omega_c p$.
We are interested in  the limit $\omega_c\to 0$ (or $\ell_B\to \infty$) and 
$p\to \infty$, while keeping the chemical potential constant,  $\omega_c p=\epsilon_F=\mu$.
To this end, we will use the following asymptotic property of Laguerre polynomials\cite{abramowitz-1964}
\begin{equation}
L_n(x)=e^{\frac{1}{2}x} J_0(2\sqrt{nx}) \; ,
\end{equation}
valid for $n\gg1$  and $x\gg1/(4n)$. Here,  $J_0$ is the  Bessel function of the first kind of order zero.
Using this property we find, 
\begin{align}
i G({\bm x},{\bm y}, t-t')&=\frac{1}{2\pi\ell_B^2}  e^{i\theta_B}  \label{eq:GBL} \\
\times\Big\{\Theta(t-t')&\sum_{m=p}^{p+\Delta} e^{-i\omega_m(t-t')} J_0[\sqrt{2m}( r/ \ell_B)]  \nonumber \\
- \Theta(t'-t)&\sum_{m=p-\Delta}^{p} e^{-i\omega_m(t-t')} J_0[\sqrt{2m}( r/ \ell_B)]\Big\} \; , 
\end{align}
with $p\gg1$, and $k_F r\gg1$. The cut-off $\Delta=(E-\omega_p)/\omega_c$ is the number of Landau levels near the Fermi energy.

To go further we use the Poisson summation formula, 
\begin{align}
\sum_{n=a}^b F(n)=\frac{F(a)+F(b)}{2}+\sum_{\ell=-\infty}^{+\infty} \int_a^b dx F(x) e^{i2\pi\ell x}\; .  \label{eq:Poisson}
 \end{align}
Then, the  propagator reads, 
\begin{align}
i G({\bm x},{\bm y}, t-t')&=\frac{1}{2\pi\ell_B^2}  e^{i\theta_B}
\nonumber\\
  \sum_{\ell=-\infty}^{+\infty}   
\Big\{\Theta(t-t')&\int_{m=p}^{p+\Delta}\!\!\!\!\!\!\!\! dm\; e^{i2\pi\ell m}e^{-i\omega_m(t-t')} J_0[\sqrt{2m}( r/ \ell_B)]  \nonumber \\
-\Theta(t'-t)&\int_{m=p-\Delta}^{p} \!\!\!\!\!\!\!\!\!\!\!\!\!dm\; e^{i2\pi\ell m} e^{-i\omega_m(t-t')} J_0[\sqrt{2m}( r/ \ell_B)]\Big\} \nonumber \\
+\frac{1}{2\pi\ell_B^2}&e^{i\theta_B}e^{-i\epsilon_F(t-t')} J_0(k_F r)\sgn(t-t') .
\end{align}
Upon the change of variables $m=(1/2) \ell_B^2 k^2$, we obtain 
\begin{align}
i G({\bm x},{\bm y}, t-t')&=\frac{1}{2\pi}  e^{i\theta_B}   \label{eq:GBLP} \\
\times  \sum_{\ell=-\infty}^{+\infty} \Big\{\Theta(t-t')&\int_{k_F}^{k_F+\lambda/2} \!\!\!\!\!\!\!\! dk k  e^{i\pi\ell \ell_B^2 k^2}e^{-i\omega_k(t-t')} J_0(k r)  \nonumber \\
- \Theta(t'-t)&\int_{k_F-\lambda/2}^{k_F}\!\!\!\!\!\!\!\!  dk k e^{i\pi\ell \ell_B^2 k^2} e^{-i\omega_k(t-t')} J_0(k r)\Big\} \nonumber \\
+\frac{1}{2\pi\ell_B^2}&e^{i\theta_B}e^{-i\epsilon_F(t-t')} J_0(k_F r)\sgn(t-t') .
\nonumber
\end{align}
We now define $q=k-k_F$ in the first integral and $q=k_F-k$ in the second one. We next use the integral representation of the Bessel function, 
\begin{equation}
J_0(x)=\int_0^{2\pi} \frac{d\varphi}{2\pi}\;e^{i x\cos\varphi}
\end{equation}
to obtain
\begin{align}
i G({\bm x},{\bm y}, t-t')&=  \label{eq:GBLPq}\\  
\frac{k_F}{2\pi} e^{-i\epsilon_F (t-t')} e^{i\theta_B}& \sum_{\ell=-\infty}^{+\infty}  \int_0^{2\pi} \frac{d\varphi}{2\pi}
 e^{ik_F r \cos\varphi}   \nonumber \\
\times\Big\{\Theta(t-t')&\int_{0}^{\lambda/2} \!\!\!\!\!\!\!\! dq  e^{iq (r\cos\varphi-v_F(t-t')+2\pi\ell \ell_B^2 k_F)}  \nonumber \\
-\Theta(t'-t)&\int_{0}^{\lambda/2} \!\!\!\!\!\!\!\! dq  e^{-iq (r\cos\varphi-v_F(t-t')+2\pi\ell \ell_B^2 k_F)}\Big\} \nonumber \\
&+ O(\ell_B^{-2}).
\nonumber
\end{align}
Implementing a smooth cut-off for the momentum integrals we finally find, 
\begin{align}
&i G({\bm x},{\bm y}, t-t')=\frac{ik_F}{2\pi} e^{-i\epsilon_F (t-t')} e^{i\theta_B}\sum_{\ell=-\infty}^{+\infty} \\ 
&\int_0^{2\pi} \frac{d\varphi}{2\pi} 
 \frac{e^{ik_F r\cos\varphi}} 
{r\cos\varphi-v_F (t-t)+2\pi\ell \ell_B^2 k_F + i\alpha\sgn(t-t)}\; , 
\nonumber
\end{align}
which coincides with Eq. \eqref{eq:Propagator}.

\section{Density of States}
\label{Ap:DensityofStates}
The density of states can be computed from the imaginary part of the self-correlation function  as, 
\begin{equation}
N(\omega)=-\sgn(\omega-\epsilon_F)\frac{1}{\pi} {\rm Im} G_F(\bm{x},\bm{x}, \omega) \; ,
\label{eq:Nomega-def}
\end{equation}
where 
\begin{align}
&G_F(\bm{x},\bm{x}, \omega)=\int_{-\infty}^{+\infty} dt\;  G_F(\bm{x},\bm{x}, t) e^{i\omega t}. 
\nonumber \\
&= \frac{k_F}{2\pi} \sum_{\ell=-\infty}^{+\infty}\int_0^{2\pi} d\varphi
\int_{-\infty}^{+\infty} dt \; G_{\ell,\varphi}(0, t) e^{-i(\epsilon_F-\omega) t} \; .
\end{align}
Notice that $N(\omega)$ is Gauge invariant since $\theta_B(\bm{x}, \bm{x})=0$ in any Gauge. 
By computing the time integral in the complex plane we find, 
\begin{align}
&G_{\ell\varphi}(\omega)=\frac{-1}{v_F}\int_{-\infty}^{+\infty}  \frac{e^{-i(\epsilon_F-\omega) t}}{t-2\pi\ell \frac{\ell_B^2k_F}{v_F}+i\alpha\sgn t}
dt
\nonumber \\
&=\frac{-2\pi i}{v_F} e^{-i2\pi\left|\ell\frac{(\omega-\omega_c)}{\omega_c} \right|}
\left[\Theta(\omega-\epsilon_F)\Theta(-\ell)-\Theta(\epsilon_F-\omega)\Theta(\ell) \right].
\end{align}
Replacing this expression into Eq. (\ref{eq:Nomega-def}) we obtain 
\begin{equation}
N(\omega-\epsilon_F)=N(0) \sum_n\delta_{n, \frac{\omega-\epsilon_F}{\omega_c}} \; , 
\label{eq:density}
\end{equation}
where $N(0)=k_F/v_F$ is the density of states at the Fermi surface of a two-dimensional Fermi gas. 
Equation (\ref{eq:density}) is just another consistency check of the asymptotic propagator, Eqs. (\ref{eq:Propagator}) and (\ref{eq:Propagator-ellvarphi}).

\section{Green function of the covariant derivative}
\label{Ap:GFD}
The covariant derivative 
\begin{equation}
\tilde D_S= \hat{\bm n}_S\cdot\bm{q}-\frac{i}{\ell_B^2k_F}\frac{\partial}{\partial \varphi_S} \; , 
\end{equation}
as well as, its inverse $D^{-1}_{S,T}(\bm{q})$
play a central role in the bosonization procedure.  For this reason, it is useful to have a deeper understanding of their properties. There are two simple, however instructive, limits: 
$\ell_B\to\infty$ (with $\bm{q}\neq 0$), and  $\bm{q}\to 0$ (with $B\neq 0$). In the former case, $\tilde D_S=\hat{\bm n}_S\cdot \bm{q}$. 
The only solution of the homogeneous equation $\tilde D_S\psi=0$, is the trivial one $\psi_0=0$. Then, given properly initial conditions, the Green function is uniquely determined and it is given by 
$D^{-1}_{S,T}=\delta_{S,T}/\hat{\bm n}_S\cdot\bm{q}$. The latter case is also quite simple. When 
$\bm{q}=0$, the periodic solution of $\tilde D_S(0)\psi=0$ is trivial, $\psi=0$,  and then, the Green function is well defined. It is given by 
\begin{equation}
D^{-1}_{ST}(0)= i \frac{\ell_B^2k_F}{2}\sum_n \sgn(\varphi_S-\varphi_T-2 n\pi) \;.
\end{equation}
However, in the presence of a finite $\bm{q}$ particle-hole pair excitation and a finite magnetic field 
$\ell_B\neq \infty$, the homogeneous equation $\tilde D_S(\bm{q})\psi=0$ has a non-trivial periodic solution,  $\psi_0(\bm{q},\varphi_S)$. Therefore, the operator $\tilde D_S(q)$ has a zero mode and, rigorously, its inverse does not exist.

Then, we can look for a modified Green function by limiting the functional space to be orthogonal to the zero mode. To be concrete, consider one wants to solve the following inhomogeneous first order linear equation, 
\begin{equation} 
\tilde D_S(\bm{q})\psi(\bm{q}, \varphi_S)= f(\bm{q}, \varphi_S)
\label{eq:inhomo}
\end{equation}
where $f(\bm{q}, \varphi_S)$ is a smooth function of $\bm{q}$ and $\varphi_S$.
Since the homogeneous equation  
\begin{equation}
\tilde D_S(\bm{q})\psi_0(\bm{q}, \varphi_S)=0 , 
\label{Ap:ZeromodeDiff}
\end{equation}
has a non-trivial solution
then, for any $f(\varphi_S)$, Eq. (\ref{eq:inhomo}) has no solution. However, it has an infinite number of solutions whether the function $f(\varphi_S)$ is orthogonal to the zero mode\cite{fredholm1903}
$\psi_0$, {\em i.\ e.\ }
\begin{equation}
\int_0^{2\pi} d\varphi_S f^*(\bm{q}, \varphi_S)\psi_0(\bm{q}, \varphi_S) =0 \;.
\end{equation}
Restricting the domain to the orthogonal functional space, the solutions of Eq. (\ref{eq:inhomo}) can be written as 
\begin{equation}
\psi(\varphi_S)=\int_0^{2\pi} d\varphi_T\;  G_m(\varphi_S, \varphi_T) f(\varphi_T) 
\end{equation}
in which we have defined  the {\em modified Green function} $G_m(\varphi_S, \varphi_T)$ satisfying 
\begin{align}
\left\{\hat{\bm n}_S\cdot\bm{q}-\frac{i}{\ell_B^2k_F} \frac{\partial}{\partial \varphi_S}\right\}& G_m(\bm{q},\varphi_S,\varphi_T)=\delta_p(\varphi_S-\varphi_T) \nonumber  \\
&-
\frac{1}{2\pi} \psi_0^*(\varphi_S)\psi_0(\varphi_T).
\label{Ap:ModifiedGreen}
\end{align}
It is immediate to show that the inhomogeneity of Eq. (\ref{Ap:ModifiedGreen}) is orthogonal to 
the zero mode $\psi_0$.

We can now solve Eq. (\ref{Ap:ModifiedGreen}) for $\varphi_S\neq \varphi_T+2n\pi$, obtaining
\begin{align}
G_m(\varphi_S,\varphi_T)&=i\ell_B^2 k_F\psi_0(\varphi_S) \times 
\nonumber \\
&\times\left\{\psi^*_0(\varphi_T)  
 -\frac{1}{2\pi}\psi_0(\varphi_T)\int_{\varphi_T}^{\varphi_S}\!\!\!\! \left[\psi_0^*(\varphi')\right]^2\; d\varphi' 
\right\}
\end{align}
(to simplify notation we are not displaying the momentum $\bm{q}$ explicitly).
We complete the calculation by imposing that 
\begin{equation}
\lim_{\varphi_T\to \varphi_S^+}G_m(\varphi_S,\varphi_T)-\lim_{\varphi_T\to \varphi_S^-}G_m(\varphi_S,\varphi_T)=i \ell_B^2 k_F
\end{equation}
The result is 
\begin{align}
G_m(\varphi_S,\varphi_T)&=\frac{i\ell_B^2 k_F}{2}\sgn(\varphi_S-\varphi_T)\psi_0(\varphi_S) \times 
\label{Ap:SolGhomo} \\
&\times\left\{\psi^*_0(\varphi_T)  
 -\frac{1}{2\pi}\psi_0(\varphi_T)\int_{\varphi_T}^{\varphi_S} \!\!\!\!\left[\psi_0^*(\varphi')\right]^2\; d\varphi' 
\right\}
\end{align}
with $\psi_0(0)=1$. By solving Eq. (\ref{Ap:ZeromodeDiff}), we see that the zero mode with the require boundary conditions in given by
\begin{equation}
\psi_0(\varphi_S,\bm{q})= e ^{-i \ell_B^2 k_F(\bm{q} \times \hat {\bm n}_S)} \; .
\label{Ap:Zeromode}
\end{equation}
The orthogonality relation with respect to the zero mode has important consequences in the form of the Green function.  In fact, we are interest in the response to reasonable smooth and isotropic electromagnetic fields. In this cases, $f(\varphi_S)\sim f =$constant. As a consequence, the inhomogeneous equation of motion will have solutions provided
\begin{equation}
\int_0^{2\pi} \psi_0(\varphi)d\varphi=2\pi J_0(q\ell^2_Bk_F)=0
\end{equation}
Thus, the momentum should be quantized in units of $1/\ell_B^2 k_F$ and are given by the zeros of the Bessel function. At low magnetic fields, the argument of the Bessel function is large and, from the asymptotic expression we can infer that the zeros are equally spaced. 
We can implemented this condition introducing the quantization in the following way
\begin{align}
G(\bm{q},\varphi_S,\varphi_T)&=\sum_n \delta_{n, q \ell_B^2k_F}G_m(\bm{q},\varphi_S,\varphi_T)
\nonumber \\
&=\sum_\ell e^{i 2\pi\ell \left(\frac{\bm{q}\cdot\bm{v}_F}{\omega_c}\right)} G_m(\bm{q},\varphi_S,\varphi_T)
\end{align} 
Finally, in order to compute $D^{-1}_{SS}$ we need to coarse grain the Green function in the patch. 
For concreteness
\begin{equation}
D^{-1}_{SS}=\int_{\varphi_S-\Lambda/2k_F}^{\varphi_S+\Lambda/2k_F}  G(\varphi_S,\varphi_T)d\varphi_T \; .
\end{equation} 
Disregarding terms of higher order in a $1/\ell_B k_F$ expansion,  we obtain, 
\begin{equation}
D^{-1}_{S,S}(q)= \sum_{\ell} \frac{e^{i 2\pi\ell \left(\bm{v}_S\cdot\bm{q}/\omega_c\right)}}{\hat{\bm n}_S\cdot\bm{q}} \; .
\label{Ap:D-1SS}
\end{equation}
We use this expression in , Eq. (\ref{eq:D-1SS}), to compute the fermion propagator. Eq. (\ref{Ap:D-1SS}) also coincides with the zero frequency limit of the diagonal Green function
\begin{equation}
D^{-1}_{SS}(\bm{q})=-\lim_{\omega-> 0} G_{SS}(\omega,\bm{q})
\end{equation}
computed in section \ref{Sec:BosDyn}.

%

\end{document}